\documentclass[final,12pt,5p,times,twocolumn,sort&compress]{elsarticle}
\usepackage{amsmath}
\usepackage{amssymb}
\usepackage{multirow}
\usepackage{bm}
\usepackage{gensymb} 
\usepackage{esint} 
\usepackage[unicode=true,colorlinks=true,linkcolor=blue,citecolor=blue,urlcolor= blue]{hyperref} 
\usepackage[english]{babel}
\usepackage[latin1]{inputenc}
\usepackage[T1]{fontenc}
\usepackage{adjustbox}
\usepackage{array}  
\usepackage{dcolumn}
\newcolumntype{L}[1]{>{\raggedright\let\newline\\\arraybackslash\hspace{0pt}}m{#1}}
\newcolumntype{C}[1]{>{\centering\let\newline\\\arraybackslash\hspace{0pt}}m{#1}}
\newcolumntype{R}[1]{>{\raggedleft\let\newline\\\arraybackslash\hspace{0pt}}m{#1}}

\newcounter{bla}

\journal{Computer Physics Communications}

\begin{document}

\begin{frontmatter}


\title{KineCluE: a Kinetic Cluster Expansion code to compute transport coefficients
		beyond the dilute limit}

\author[a,b]{Thomas Schuler\corref{cor1}}
\author[a]{Luca Messina}
\author[a]{Maylise Nastar}

\cortext[cor1] {Corresponding author.\\\textit{E-mail address:} thomas.schuler@cea.fr}
\address[a]{DEN-Service de Recherches de M\'etallurgie Physique, CEA, Universit\'e Paris-Saclay, F-91191 Gif-sur-Yvette, France}
\address[b]{Mines Saint-Etienne, Universit\'e de Lyon, CNRS, UMR 5307 LGF, Centre SMS, F-42023 Saint-Etienne, France}

\begin{abstract}
This paper introduces the KineCluE code that implements the self-consistent mean-field theory for clusters of finite size. Transport coefficients are obtained as a sum over cluster contributions (in a cluster expansion formalism), each being individually computed with KineCluE. This method allows for the calculation of these coefficients beyond the infinitely dilute limit, and is	an important step in bridging the gap between dilute and concentrated approaches. Inside a finite volume of space containing the components	of a given cluster, all kinetic trajectories are accounted for in an exact manner. The code, written in Python, adapts to a wide variety of systems, with various crystallographic structures (possibly under strain), defects and solute amount and types, and various jump mechanisms, including collective ones. The code also features a set of useful tools, such as the sensitivity study routine that allows for the identification of the most important jump frequencies to get accurate transport coefficients with minimum computational cost.
\end{abstract}


\end{frontmatter}

{\bf PROGRAM SUMMARY}

\begin{small}
\noindent
\\
{\em Program Title: KineCluE (KINEtic CLUster Expansion}) \\
{\em Licensing provisions: LGPL} \\
{\em Programming language: Python 3.6} \\
\\
{\em Nature of problem: Providing a general method for computing transport coefficients from atomic jump frequencies, taking into account kinetic correlations.} \\
\\
{\em Solution method: The program relies on the self-consistent mean field (SCMF) theory. The system is described in terms of lattice sites, defects and jump mechanisms. The first part of the code translates the diffusion problem for such system into an analytical linear eigenvalue problem. The second  part of the code assigns numerical values to each analytical variable and then solves the linear problem.}\\
\end{small}


	\section{Introduction}
	Atomic transport in solids has attracted enormous experimental \cite{LBdiffusion} and theoretical \cite{leclaire:1956,allnatt:1993} attention over the past 70 years. It is still a challenging problem in various ways. In the first place, the microscopic dynamical rules determining the macroscopic diffusion coefficients are becoming more and more complex as the investigations based on density functional theory (DFT) \cite{martin:2004} provide an increasingly detailed description of the atomic transport mechanisms \cite{meslin:2007,messina:2014,Barkema1996}. In addition, the effort of solving the $n$-body problem of $n$ interacting species diffusing on a lattice has been mainly focused on the limiting case of two-body diffusion problems both in dilute \cite{allnatt:1993} and -- with simplifying assumptions -- in concentrated alloys \cite{kikuchi:1961,kikuchi:1970,nastar:2006b, barbe:2007b,ERDELYI2016,Vaks2016}. Three-body diffusion problems have only been tackled in very simple systems \cite{barbe:2006,bocquet:2015}. Besides, the experimental characterization of atomic transport at low temperature is in general not feasible, and only some of the transport coefficients can be experimentally determined. Thanks to the progress of DFT applied to the calculation of migration barriers \cite{mills:1995, jonsson:1998, henkelman:2000}, atom jump-frequency databases are in construction for metallic alloys \cite{mantina:2009,choudhury:2011, messina:2014, wu:2017, agarwal:2018}, and to a smaller extent for ceramics \cite{schulerfuel:2016, schulerfuel:2017} and semiconductors \cite{yang:2003,caliste:2007,bruneval:2012,yoshida:2016,kobayashi:2017}. The use of these databases is mainly dedicated to the study of self- and solute diffusion \cite{mantina:2009,wu:2017}. Much less attention has been paid to the estimation of the full Onsager matrix \cite{onsager:1931}, although the latter is essential to investigate flux coupling between point defects (PD) and atomic species occurring in systems driven by a PD supersaturation, in materials subjected to irradiation \cite{nastar:2013,Ardell2016} as well as during thermal quenching treatments \cite{anthony:1975} or under severe plastic deformation \cite{zehetbauer:2006}.

	In most cases, diffusion mechanisms in solids are mediated by PDs, the most frequent ones being the exchange of an atom with a first nearest neighbor (1NN) vacancy (i.e., a vacant lattice site) and the exchange of an atom with a 1NN self interstitial. One of the most stable interstitial configuration is the dumbbell (or split-interstitial) configuration: a directional pair of atoms sharing a lattice site. This is a complex mechanism possibly leading to a modification of the initial dumbbell composition and direction, and there is still no exact calculation of the associated transport coefficients even in the limiting case of an infinite dilute alloy (a single PD and a solute atom pair in a host matrix) \cite{barbu:1980, bocquet:1991, barbe:2006c, barbe:2007b}.
	Due to the small PD concentration, the successive jumps of a given atom are correlated because the probability of making several exchanges with the same PD is larger than the probability of making exchanges with a different one. These kinetic correlations, related to the probability for the PD to perform return paths, slow down the diffusion of a tracer atom and lead to non-zero off-diagonal Onsager coefficients $L_{AB}$ between two atomic species $A$ and $B$. Thus, they determine the sign and the amplitude of flux coupling. In the limiting case of an infinitely dilute alloy containing a vacancy, automated computational schemes have allowed for the calculation of transport coefficients for long-range thermodynamic interactions between a solute and a PD \cite{PhysRevB.88.134201, PhysRevB.89.144202, messina:2014,schuler:2016, schuler:2017,agarwal:2017,agarwal:2018,bocquet:2014,trinkle:2017}. At the origin of these numerical schemes are new theoretical developments based either on the Self-Consistent Mean Field theory (SCMF)  \cite{schuler:2016} or on a Green-function formalism \cite{trinkle:2016, trinkle:2017,bocquet:2014,bocquet:2015}. Indeed, even though computing a macroscopic transport coefficient from microscopic jump rates is a well-defined problem, many efforts are still needed to find the general solution.
	
	Within a diffusion theory, the evolution of an alloy is described by means of a microscopic master equation. One starts from a simplified description of the system as a rigid lattice, whose sites are occupied by either atoms or PDs. Transitions from one configuration to another are fixed by well-defined jump mechanisms. These transitions determine the evolution of the distribution function, e.g., the probability of every configuration as a function of time. Solving the master equation and computing the time-dependent distribution function entails finding a non-linear eigenvalue problem coupled to the unknown diffusion driving forces, i.e., the gradients of chemical potentials. At first order in the diffusion driving forces and in stationary conditions, the problem becomes a linear eigenvalue problem, where the eigenvalues correspond to the contribution of kinetic correlations to the Onsager coefficients. 
	An exact solution has been obtained so far only in the particular case of a vacancy-solute cluster in an infinite medium, thanks to a Green-function formalism used to solve kinetic equations in close relation to the SCMF formalism \cite{trinkle:2017}. When there is no exact method, we can use the exact variational formula demonstrating that phenomenological coefficients correspond to a minimum of a functional \cite{kipnis:1999, spohn:1991}. It is thus always possible to derive approximate methods that provide upper bounds of the phenomenological coefficients and progressively converge towards the exact solution \cite{arita:2018}.
	Within the SCMF theory, kinetic correlations are represented by a non-equilibrium effective Hamiltonian \cite{nastar:2000}, and are assumed to be a linear combination of the chemical potential gradients, leading to a solution in the form of partial differential kinetic equations \cite{nastar:2014}. A solute-PD $n$-body kinetic interaction  is related to the kinetic equation of the corresponding $n$-th moment of the probability distribution function \cite{nastar:2000,nastar:2005}. In a dilute alloy, the  $n$-th moment is assimilated to the ensemble probabilities of the various configurations of the solute-PD cluster of size $n$ \cite{schuler:2016}. The range of a kinetic interaction is directly related to the length of the return paths of the cluster components. For example, a 1NN pairwise kinetic interaction accounts only for the two-jump return paths of a PD. A systematic increase of the kinetic interaction range yields the Onsager coefficients as a series converging to the exact Onsager solution of a single cluster in an infinite medium. Note that in most cases, modeling a cluster in a finite medium is closer to reality because at some point, a migrating PD may be trapped at structural defects or in the vicinity of solute atoms. A solution in the form of a series whose successive terms are associated with increasingly long migration paths of the defect provides a powerful method of studying finite-size effects on kinetic properties.
The use of symmetry operations for the analysis of the point-defect exchange frequencies and the computation of the Onsager coefficients have recently widen the investigations to more complex crystallographic structures, as for example the study of solute drag in strained systems \cite{PhysRevB.88.134108} and in the hexagonal close-packed magnesium crystal \cite{agarwal:2017}.  However, we still miss a general kinetic method that can handle complex diffusion mechanisms, such as for instance the dumbbell diffusion mechanism \cite{bocquet:1991}, the kick-out mechanism  \cite{Fu:2005b}, or the macro-jumps of two-half vacancies surrounding a large-sized solute atom  \cite{bocquet:2017}. Moreover, a systematic method dealing with clusters larger than two components does not yet exist. In these cases, the challenge is to achieve a full exploration of the configuration phase space, reduce it with the aid of symmetry operations, then build a graph whose vertices represent the configurations and edges represent the authorized transitions. 
	
	In this work, we address all of the above issues with KineCluE, an automated code aimed at the computation of transport coefficients in dilute alloys with intermediate solute concentrations between the infinitely-dilute and the concentrated limits. Recently, we have derived an approximation scheme based on a division of the system into PD-solute cluster sub-spaces \cite{schuler:2016,EqLoc}. Computing separately the transport coefficient contribution of each cluster, and then modeling the system as a lattice gas of clusters leads to a very efficient approach of diffusion and phase transformations in dilute systems. Within this new formalism, the total Onsager matrix is split into intrinsic cluster Onsager matrices, allowing as well for a proper definition of the cluster mobility and dissociation rate. The code is versatile and able to deal with many types of crystal structures, defect types, and jump mechanisms, including collective ones. Its ability to perform calculations on clusters larger than pairs is an important step in bridging the gap between dilute and concentrated approaches to diffusion problems. 
    
    The paper is structured as follows. Section \ref{sec:theory} presents the theoretical framework behind KineCluE. The technical implementation of the code is described in Sec. \ref{sec:techno}.  Then, some interesting features are detailed and the code performance is assessed in Sec. \ref{sec:perf}. Finally, application examples and validation tests are presented in Sec. \ref{sec:tests}.

	\section{Theoretical background \label{sec:theory}}
	
	\subsection{Cluster transport coefficients\label{sec:cluster}}
	Transport coefficients appear in the framework of thermodynamics
	of irreversible processes, originally developed by Onsager \cite{onsager:1931,onsager:1931bis}:
	
	\begin{equation}
	\vec{J}_{\alpha}=-\sum_{\beta}L_{\alpha\beta}\vec{\nabla}\frac{\mu_{\beta}}{k_\mathrm{B}T},\label{eq:tip}
	\end{equation}    
	where $L_{\alpha\beta}$ is the Onsager transport coefficient (units of $\left[\mathrm{m^{-1}s^{-1}}\right]$)
	relating the flux $\vec{J}_{\alpha}$ of species $\alpha$ under a
	chemical potential gradient $\vec{\nabla}\mu_{\beta}$ of species
	$\beta$, $k_\mathrm{B}$ is the Boltzmann constant, and $T$ is the absolute
	temperature. The Allnatt formula relates the transport coefficients
	with equilibrium fluctuations of atomic positions \cite{Allnatt_1982}:
	\begin{equation}
	L_{\alpha\beta}=\lim_{\tau\rightarrow\infty}\frac{\left\langle \Delta\vec{R_{\alpha}}\left(\tau\right)\Delta\vec{R_{\beta}}\left(\tau\right)\right\rangle }{6V\tau},\label{eq:allnatt}
	\end{equation}    
	where $V$ is the total volume of the system and $\Delta\vec{R_{\alpha}}\left(\tau\right)$
	the total displacement vector of atoms of species $\alpha$ during
	time-step $\tau$. Starting from this definition, we define \textit{cluster transport coefficients}, under the assumption that
	the system is sufficiently dilute so that it can be divided
	into kinetically independent sub-spaces called clusters \cite{schuler:2016,schuler:2017,EqLoc}. These coefficients are intrinsic equilibrium
	properties of each cluster, and the total transport coefficients are
	obtained from the relation:
	
	\begin{equation}
L_{\alpha\beta}=\sum_{c}\left[c\right]^\mathrm{oe}L_{\alpha\beta}^\mathrm{eq}\left(c\right),\label{eq:clustercoeff}
	\end{equation}
	where $L_{\alpha\beta}^\mathrm{eq}\left(c\right)$ is the transport coefficient
	of cluster $c$ (units of $\left[\mathrm{m^{2}s^{-1}}\right]$) ,
	and $\left[c\right]^\mathrm{oe}$ is the concentration of cluster
	$c$ per unit volume, which is not necessarily an equilibrium concentration. Thus, Eq. \ref{eq:clustercoeff}
	defines out-of-equilibrium macroscopic transport coefficients, and expanding these
	coefficients into larger clusters enables the model to go beyond the dilute
	limit. Cluster transport coefficients defined this way have the same
	units as diffusion coefficients. In order to have the same units
	between cluster- and total transport coefficients,
	the former must be divided by the atomic volume, in which case the concentrations
	in Eq. \ref{eq:clustercoeff} become site concentrations.
	
	KineCluE computes cluster transport coefficients as functions of
	temperature and strain, using
	self-consistent mean-field theory \cite{nastar:2000,nastar:2005}. Since a change in the chemical potentials affects only the cluster concentrations in Eq. \ref{eq:clustercoeff}, KineCluE
	is a crucial step in developing a general and efficient framework to provide accurate atomic-based kinetic properties for higher-scale models such as object kinetic Monte Carlo \cite{Vattre2016,messinab:2016,Wirth2017,Castin2018,CHIAPETTO2017,Adjanor15,Stoller12}, cluster dynamics \cite{Jourdan2014,Stoller08,Soisson2016,Clouet_ASM} or phase-field models \cite{Badillo2015,piochaud:2016,thuinet:2018}.

	\subsection{Self-consistent mean-field theory}
	
	The goal of the self-consistent mean-field (SCMF) theory is to compute
	transport coefficients as a thermodynamic -- i.e., equilibrium -- average of atomic jumps.
	A chemical potential gradient (CPG) driving the system out of equilibrium
	is assumed, and the resulting flux is computed using a thermodynamic
	average. Transport coefficients are then identified from Eq.
	\ref{eq:tip}. The method has been described in details elsewhere \cite{nastar:2005,schuler:2016}, hence we only outline here the main steps to obtain an alternative formulation, more suitable for coding.
	
	It is assumed that the system can be mapped onto a rigid lattice containing
	a number of lattice sites, each being occupied by a single atom or
	defect. The microscopic master equation controls the evolution of
	a system represented by a configuration vector $\mathbf{n}$, whose
	components are the site occupation numbers $n_{i}^{\alpha}$ ($n_{i}^{\alpha}=1$
	if species $\alpha$ occupies site $i$ in configuration $\mathbf{n}$,
	and $n_{i}^{\alpha}=0$ if not):
	
	\begin{equation}
	\frac{dP\left(\mathbf{n},t\right)}{dt}=\sum_{\tilde{\mathbf{n}}}\left[W\left(\tilde{\mathbf{n}},\mathbf{n}\right)P\left(\tilde{\mathbf{n}},t\right)-W\left(\mathbf{n},\tilde{\mathbf{n}}\right)P\left(\mathbf{n},t\right)\right].\label{eq:masterEQ}
	\end{equation}
	$P\left(\mathbf{n},t\right)$ is the probability of having configuration
	$\mathbf{n}$ at time $t$, and $W\left(\tilde{\mathbf{n}},\mathbf{n}\right)$
	is the rate at which a system in configuration $\tilde{\mathbf{n}}$
	transitions to configuration $\mathbf{n}$. It is assumed that the probability
	of any configuration can be expressed as the product of its equilibrium
	probability $P_{0}\left(\mathbf{n}\right)$ and a probability $\delta P\left(\mathbf{n},t\right)$
	that corresponds to the deviation from equilibrium $P\left(\mathbf{n},t\right)=P_{0}\left(\mathbf{n}\right)\, \delta P\left(\mathbf{n},t\right)$,
	and that $\delta P\left(\mathbf{n},t\right)$ has the same mathematical
	form as the equilibrium probability. However, in $\delta P\left(\mathbf{n},t\right)$ thermodynamic interactions are replaced by an effective Hamiltonian, to account for the fact
	that two equivalent configurations at equilibrium do not necessarily
	have the same probability in out-of-equilibrium conditions, because
	the driving force breaks the symmetry of the system. In our formulation of the SCMF theory,
	the effective Hamiltonian is reduced to $n_{c}$-body effective interactions,
	where $n_{c}$ is the number of components (defects, solutes) in cluster
	$c$. The introduction of a driving force (the CPG) reduces the symmetry of
	the system such that only the crystal symmetry operations conserving
	the CPG direction are valid for the out-of-equilibrium system. We group all symmetry-equivalent (in the out-of-equilibrium system) configurations into effective interaction classes because all of such configurations will equally contribute to the non-equilibrium averages. The magnitude of effective interactions belonging to class $\sigma$ is denoted $\nu_{\sigma}$, and $n_{\sigma}$ represents a product of site occupancies for each of the $n_{c}$ cluster components, times a sign variable. Hence, for a given configuration, $n_{\sigma}=\pm 1$ if this configuration is identical to one of the instances of class $\sigma$, and 0 if it is not.
	Knowing the sign variable for one instance of the effective interaction class, the sign variable of another instance is the same if the symmetry operation transforming one instance
	into the other maintains the CPG vector, while the sign variable is the opposite if the symmetry operation transforms the CPG vector into its opposite. Therefore, the deviation from the equilibrium probability is written as:
	
	\begin{align}
	\delta P\left(\mathbf{n},t\right)=\exp\left(\delta\Omega+\sum_{i,\alpha}n_{i}^{\alpha}\dfrac{\delta\mu_{i}^{\alpha}}{k_{B}T}-\sum_{\sigma}n_{\sigma}\dfrac{\nu_{\sigma}}{k_\mathrm{B}T}\right),\label{eq:deltaP}
	\end{align}
	where $\delta\mu_{i}^{\alpha}$ is the local (on site $i$) deviation from
	the equilibrium chemical potential of species $\alpha$ and $\delta\Omega$
	is a normalization constant. The quantities $\delta\Omega$, $\delta\mu_{i}^{\alpha}$, and
	$\nu_{\sigma}$ are time dependent, but since we are only interested in
	the steady-state flux, the time dependence is omitted for simplicity.
	Note that using only $n_{c}$-body effective
	interactions to describe the deviation from equilibrium is not restrictive because it
	fully characterizes a system of $n_{c}$-conserving defects and solutes in a
	bulk matrix. 
	
	The continuity equation per site reads:
	
	\begin{equation}
	\frac{d\left[\alpha\right]_{i}}{dt}=-\oiint_{S}\vec{J}_{i}^{\alpha}.d\vec{S}=-\sum_{s\in\theta_{i}^{\alpha}}\Gamma_{i\rightarrow s}^{\alpha},\label{eq:continuity}
	\end{equation}
	where $\left[\alpha\right]_{i}$ is the probability of site $i$ to
	be occupied by species $\alpha$ (hence the local site concentration),
	$J_{i}^{\alpha}$ is the local flux per unit surface of species $\alpha$
	from site $i$, and $\Gamma_{i\rightarrow s}^{\alpha}$ is the
	rate at which atoms of species $\alpha$ jump from site $i$ to site
	$s$, the latter being located at jumping distance from site $i$
	($s\in\theta_{i}^{\alpha}$). The site concentration of species
	$\alpha$ on site $i$ is also given by the first moment of the probability
	distribution function:
	
	\begin{equation}
	\left[\alpha\right]_{i}=\left\langle n_{i}^{\alpha}\right\rangle ^\mathrm{oe}=\sum_{\mathbf{n}}n_{i}^{\alpha}P\left(\mathbf{n},t\right).\label{eq:1stmoment}
	\end{equation}
	$\left\langle\cdot\right\rangle ^\mathrm{oe}$ denotes the ensemble average
	over the out-of-equilibrium distribution function $P\left(\mathbf{n},t\right)$.
	We combine Eqs. \ref{eq:masterEQ}-\ref{eq:1stmoment} to obtain the
	atomic-scale description of the variation of the local concentration over time:
	
	\begin{align}
	& \frac{d\left\langle n_{j}^{\beta}\right\rangle^\mathrm{oe} }{dt}=\left\langle \sum_{\mathbf{\tilde{n}}}n_{j}^{\beta}W\left(\mathbf{n},\tilde{\mathbf{n}}\right)\left[\delta P\left(\mathbf{\tilde{n}}\right)-\delta P\left(\mathbf{n}\right)\right]\right\rangle \nonumber \\
	& =\left\langle \sum_{\mathbf{\tilde{n}}}n_{j}^{\beta}W\left(\mathbf{n},\tilde{\mathbf{n}}\right)\left[\sum_{i,\alpha}\left(\tilde{n}_{i}^{\alpha}-n_{i}^{\alpha}\right)\dfrac{\delta\mu_{i}^{\alpha}}{k_\mathrm{B}T}\right.\right.\nonumber \\
	& \quad\left.\left.-\sum_{\sigma}\left(\tilde{n}_{\sigma}-n_{\sigma}\right)\dfrac{\nu_{\sigma}}{k_\mathrm{B}T}\right]\right\rangle .\label{eq:scmf_step1}
	\end{align}
	The first equality makes use of the detailed balance principle $W\left(\tilde{\mathbf{n}},\mathbf{n}\right)P_{0}\left(\tilde{\mathbf{n}},t\right)=W\left(\mathbf{n},\tilde{\mathbf{n}}\right)P_{0}\left(\mathbf{n},t\right)$,
	and $\left\langle\cdot\right\rangle $ denotes the ensemble average over
	the equilibrium distribution function $P_{0}\left(\mathbf{n}\right)$,  implying a sum over all configurations \textbf{n}.
	The second equality arises from a first-order expansion of the exponential
	function in Eq. \ref{eq:deltaP}. The expression in Eq. \ref{eq:scmf_step1}
	can be greatly simplified for the following reasons.
	\begin{enumerate}
	\item For transitions between two configurations $\tilde{\mathbf{n}}$ and $\mathbf{n}$ where atom  $\beta$ at site $j$ does not move, all terms in the bracket
	will cancel out when $\tilde{\mathbf{n}}$ and $\mathbf{n}$ are inverted
	in the double sum over all configurations (one sum is
	written explicitly, while the other is implicit in the $\left\langle\cdot\right\rangle $
	symbol). Thus the sum over $\tilde{\mathbf{n}}$ is restricted to
	configurations where $\beta$ is one jump away from site $j$, that
	is the ensemble of sites $s\in\theta_{j}^{\beta}$. Note that
	additional constraints might exist to make a jump possible, for instance the fact that
	a substitutional atom needs a vacancy on the destination site. These
	constraints, expressed as a product of site occupation numbers, are
	denoted as $m_{js}^{\beta}$, and the rate  of
	such jump is $\omega_{js}^{\beta}$ $\left[s^{-1}\right]$;
	\item All sites in the system are occupied either by a "bulk-like" species or by a solute or defect.
	Only chemical potential differences thus appear: $\delta\bar{\mu}_{i}^{\alpha}=\delta\mu_{i}^{\alpha}-\delta\mu_{i}^{\mathrm{bulk}}$,
	and the sum over species $\alpha$ is restricted to defects and solutes
	belonging to the cluster; 
	\item The driving force is assumed to be homogeneous in the system: $\delta\bar{\mu}_{k}^{\alpha}-\delta\bar{\mu}_{i}^{\alpha}=\vec{ik}\cdot\vec{\nabla}\bar{\mu}_{\alpha}=d_{ik}^{\mu}\nabla\bar{\mu}_{\alpha}$,
	where $k$ is the location of species $\alpha$ after the jump and
	$d_{ik}^{\mu}$ is the jump distance projected along the unit vector $\vec{e}_{\mu}$ that is collinear with the CPG direction ($d_{ik}^{\mu}$ may be positive or negative).
	\end{enumerate}
    Therefore, Eq. \ref{eq:scmf_step1} becomes:
	
	\begin{align}
	& \left\langle \sum_{s\in\theta_{j}^{\beta}}n_{j}^{\beta}m_{js}^{\beta}\omega_{js}^{\beta}\left[\sum_{\alpha}\sum_{i,k}n_{i}^{\alpha}\tilde{n}_{k}^{\alpha}d_{ik}^{\mu}\dfrac{\nabla\bar{\mu}_{\alpha}}{k_\mathrm{B}T}\right.\right.\nonumber \\
	& \quad\left.\left.-\sum_{\sigma}\left(\tilde{n}_{\sigma}-n_{\sigma}\right)\dfrac{\nu_{\sigma}}{k_\mathrm{B}T}\right]\right\rangle =-\sum_{s\in\theta_{j}^{\beta}}\Gamma_{j\rightarrow s}^{\beta}.\label{eq:scmf_step2}
	\end{align}
	The flux from site $j$ along a particular diffusion direction $\vec{e}_{d}$ (unit vector) is obtained by summing $\Gamma_{j\rightarrow s}^{\beta}$ over all forward jumps weighted by ${d_{js}^{d}/V_\mathrm{at}=\overrightarrow{js}\cdot\vec{e}_{d}/V_\mathrm{at}}$, where $V_\mathrm{at}$ is the volume per site. Ensemble $\theta_{j+}^{\beta}$ indicates that the summation runs over jumps with a positive $d_{js}^{d}$ only. The macroscopic flux along $\vec{e}_{d}$ in a system of volume $V$ is then obtained as an average over all sites in the system: 
	
	\begin{equation}
	\mathbf{J}_{d,\beta}=\vec{J}_{\beta}.\vec{e}_{d}=\frac{1}{V}\sum_{j}n_{j}^{\beta}\sum_{s\in\theta_{j+}^{\beta}}d_{js}^{d}\Gamma_{j\rightarrow s}^{\beta}.\label{eq:macroflux}
	\end{equation}
	Flux $\Gamma_{j\rightarrow s}^{\beta}$ is identified from Eq.
	\ref{eq:scmf_step2} for each specific jump, and
	we use for convenience a matrix notation: $\mathbf{J}_{d}=-\frac{1}{V}\left(\boldsymbol{\Lambda}^{0}_{d}\boldsymbol{\mu}-\boldsymbol{\Lambda}_{d}\boldsymbol{\nu}\right)$,
	where $\boldsymbol{\mu}$ and $\boldsymbol{\nu}$ are
	vectors that do not depend on the diffusion direction but rather on the CPG direction. $\boldsymbol{\mu}$ is a vector of length $N_\mathrm{spec}$ whose components are $\nabla\bar{\mu}_{\alpha}/k_\mathrm{B}T$ ($N_\mathrm{spec}$ being the number of species); $\boldsymbol{\nu}$ is a vector of length $N_\mathrm{inter}$ whose components are the values of the effective interactions 
	$\nu_{\sigma}/k_\mathrm{B}T$ ($N_\mathrm{inter}$ being the number of effective interactions); $\mathbf{J}_{d}$ is a vector of length $N_\mathrm{spec}$ whose components are $\mathbf{J}_{d,\beta}$, the fluxes of each species $\beta$ in the diffusion direction $\vec{e}_{d}$; $\boldsymbol{\Lambda}_{d}^{0}$ is a matrix of size $(N_\mathrm{spec}\, ,\, N_\mathrm{spec})$ representing the uncorrelated contribution to diffusion in direction $\vec{e}_{d}$, while the correlated contribution in the same direction is contained in $\boldsymbol{\Lambda}_{d}$, a matrix of size $(N_\mathrm{spec}\, ,\, N_\mathrm{inter})$. The components of these two matrices are given hereafter:

	\begin{equation}
	\boldsymbol{\Lambda}_{d,\beta\alpha}^{0}=\sum_{j}\sum_{s\in\theta_{j+}^{\beta}}d_{js}^{d}\left\langle n_{j}^{\beta}m_{js}^{\beta}\omega_{js}^{\beta}\sum_{i,k}n_{i}^{\alpha}\tilde{n}_{k}^{\alpha}d_{ik}^{\mu}\right\rangle ,\label{eq:lambda0}
	\end{equation}
	
	\begin{equation}
	\boldsymbol{\Lambda}_{d,\beta\sigma}=\sum_{j}\sum_{s\in\theta_{j+}^{\beta}}d_{js}^{d}\left\langle n_{j}^{\beta}m_{js}^{\beta}\omega_{js}^{\beta}\left(n_{\sigma}-\tilde{n}_{\sigma}\right)\right\rangle .\label{eq:lambda}
	\end{equation}
	
	Values for the effective interactions are obtained from the stationarity
	of the $n_{c}{}^\mathrm{th}$-moment equations $\frac{d}{dt}\left\langle n_{i_{1}}^{\alpha_{1}}n_{i_{2}}^{\alpha_{2}}..n_{i_{n}}^{\alpha_{n}}\right\rangle^{oe} =\frac{d}{dt}\left\langle n_{\sigma_{0}}\right\rangle^{oe}=0$
	for each configuration of the system corresponding to one instance
	$\sigma_{0}$ of some effective interaction class $\sigma$: 
	
	\begin{align}
	& \quad\left\langle n_{\sigma_{0}}\sum_{j,\alpha}n_{j}^{\alpha}\sum_{s\in\theta_{j}^{\alpha}}m_{js}^{\alpha}\omega_{js}^{\alpha}d_{js}^{\mu}\nabla\bar{\mu}_{\alpha}\right\rangle \nonumber \\
	& =\left\langle n_{\sigma_{0}}\sum_{j,\alpha}n_{j}^{\alpha}\sum_{s\in\theta_{j}^{\alpha}}m_{js}^{\alpha}\omega_{js}^{\alpha}\left(-\nu_{\sigma_{0}}+\sum_{\sigma}\tilde{n}_{\sigma}\nu_{\sigma}\right)\right\rangle .\label{eq:scmf_step4}
	\end{align}
	
	Note that the thermodynamic average $\left\langle .\right\rangle $
	reduces to one configuration only for each effective interaction $\sigma_{0}$.
	Also, the sum over site $j$ and species $\alpha$ appears because
	any component of the $n_{c}$-component cluster can potentially move
	and create non-zero contributions to both sides of the equation. Equation
	\ref{eq:scmf_step4} is more conveniently written in matrix format:
	\begin{equation}
	    \mathbf{M}\boldsymbol{\mu}=\mathbf{T}\boldsymbol{\nu}, \label{eq:systemnosite}
	\end{equation}
	with
	
	\begin{equation}
	\mathbf{M}_{\sigma_{0}\alpha}=\left\langle n_{\sigma_{0}}\sum_{j}n_{j}^{\alpha}\sum_{s\in\theta_{j}^{\alpha}}m_{js}^{\alpha}\omega_{js}^{\alpha}d_{js}^{\mu}\right\rangle ,\label{eq:M}
	\end{equation}
	\begin{equation}
	\mathbf{T}_{\sigma_{0}\sigma}=\left\langle n_{\sigma_{0}}\sum_{j,\alpha}n_{j}^{\alpha}\sum_{s\in\theta_{j}^{\alpha}}m_{js}^{\alpha}\omega_{js}^{\alpha}\left(\tilde{n}_{\sigma}-\delta_{\sigma_{0}\sigma}\right)\right\rangle .\label{eq:T}
	\end{equation}
and $\delta_{\sigma_{0}\sigma}$  is the Kronecker symbol for effective interactions $\sigma_{0}$ and $\sigma$.
	
	Then, solving for effective interactions consists in inverting matrix
	$\mathbf{T}$, thus obtaining $\boldsymbol{\nu}=\mathbf{T}^{-1}\mathbf{M}\boldsymbol{\mu}$.
	This solution is inserted in the flux expression: $\mathbf{J}_{d}=-\frac{1}{V}\left(\boldsymbol{\Lambda}^{0}_{d}\boldsymbol{\mu}-\boldsymbol{\Lambda}_{d}\boldsymbol{\nu}\right)=-\frac{1}{V}\left(\boldsymbol{\Lambda}^{0}_{d}-\boldsymbol{\Lambda}_{d}\mathbf{\mathbf{T}^{-1}\mathbf{M}}\right)\boldsymbol{\mu}$
	from which the cluster transport coefficients are identified in the diffusion
	direction $\vec{e}_{d}$: $L_{d,\beta\alpha}^{eq}\left(c\right)=\frac{1}{V}\left(\boldsymbol{\Lambda}^{0}_{d}-\boldsymbol{\Lambda}_{d}\mathbf{\mathbf{T}^{-1}\mathbf{M}}\right)_{\beta\alpha}$. The next section shows that $\mathbf{M}$ is proportional to  $\boldsymbol{\Lambda}_{\mu}^{t}$, which allows for a reduction of computational loads and memory needs.
		
	\subsection{Relationship between M and $\Lambda$ \label{sec:Mlambda}}
	
	Since the expressions for $\mathbf{M}$ and $\boldsymbol{\Lambda}$ present
	interesting similarities (Eqs. \ref{eq:lambda} and \ref{eq:M}), we show how to obtain $\mathbf{M}$ directly 
	from $\boldsymbol{\Lambda}$, without the need for additional calculations.
    
    The complete derivation is provided in \ref{MLambda_anx}, and amounts to the following equation:
    \begin{equation}
	N_{\sigma}\mathbf{M}_{\sigma_{0}\alpha}=\boldsymbol{\Lambda}_{\mu,\alpha\sigma},\label{eq:EQ_M_LambdaMain}
	\end{equation}
	where $N_{\sigma}$ -- the number of symmetry equivalents for effective interaction $\sigma$ -- is easily obtained from symmetry operations for the out-of-equilibrium crystal. The components of the $\mathbf{M}$ matrix do not depend on the diffusion direction, but rather on the CPG direction $\vec{e}_{\mu}$. Therefore, as long as $\vec{e}_{\mu}$ is taken as one of the diffusion directions (which is always the case in KineCluE), the computation of $\mathbf{M}$ can be avoided.
	
	Remarkably, this relation (Eq. \ref{eq:EQ_M_LambdaMain}) guarantees that the cluster transport-coefficient matrix is symmetric, at least for $d=\mu$. In order to prove this, let us define a diagonal matrix $\mathbf{N}$ which contains the $N_{\sigma}$ coefficients. Then $\boldsymbol{\Lambda}^{t}_{\mu}=\mathbf{N}\mathbf{M}$
	and the cluster transport coefficient matrix is expressed as:
	
	\begin{equation}
	\mathbf{L}^\mathrm{eq}_{d}\left(c\right)=\frac{\boldsymbol{\Lambda}^{0}_{d}-\boldsymbol{\Lambda}_{d}\mathbf{\mathbf{T}^{-1}\mathbf{N}^{-1}}\boldsymbol{\Lambda}^{t}_{\mu}}{V}=\frac{\boldsymbol{\Lambda}^{0}_{d}-\boldsymbol{\Lambda}_{d}\mathbf{\tilde{\mathbf{T}}^{-1}}\boldsymbol{\Lambda}^{t}_{\mu}}{V}, \label{eq:matrixnotation}
	\end{equation}
	with $\tilde{\mathbf{T}}=\mathbf{NT}$. Since 
	$\tilde{\mathbf{T}}$ is symmetric because of the detailed balance, $\mathbf{L}_{\mu}^\mathrm{eq}\left(c\right)$ is also symmetric.
    
    \subsection{Accounting for site interactions \label{sec:siteinter}}
	
    We have assumed so far that a system of $n_c$ components is fully characterized by $n_c$-body effective interactions, such that neglecting all effective interactions involving a smaller number of components (let us call them \textit{sub-cluster interactions}) is correct in the sense that the cluster transport coefficients will not vary whether these sub-cluster interactions are accounted for or not. However, this is not always true because the sub-space of the configuration space defined here as the cluster is of finite size. We show hereafter that a rigorous treatment of the boundary between \textit{associated}  (i.e., inside the cluster) and \textit{dissociated} configurations (i.e., outside the cluster) requires sub-cluster interactions to be explicitly taken into account in some systems.

With Eq. \ref{eq:scmf_step4}, we ensured that $\frac{d}{dt}\left\langle n_{i_{1}}^{\alpha_{1}}\dots n_{i_{n}}^{\alpha_{n}}\right\rangle^\mathrm{oe}=0$ as long as the configuration containing atomic species $\alpha_{1}\dots \alpha_{n}$ at sites $i_{1}\dots i_{n}$ is associated. The question is whether or not this is sufficient to ensure that any sub-cluster probability is stationary.
    
    \begin{equation}
    	\frac{d}{dt}\left\langle n_{i_{1}}^{\alpha_{1}}\dots n_{i_{n-1}}^{\alpha_{n-1}}\right\rangle^\mathrm{oe}=\sum_{i_{n}\neq i_{1}\dots i_{n-1}} \frac{d}{dt}\left\langle n_{i_{1}}^{\alpha_{1}}\dots n_{i_{n}}^{\alpha_{n}}\right\rangle^\mathrm{oe},\label{eq:msite0}
    \end{equation}
    where the sum runs over all possible sites for $i_n$. Two types of sites can be distinguished for $i_n$: sites where the $n_c$ components form an associated configuration, and those where they form a dissociated one. The sum over the first type of sites is null because each term is null, as is ensured by Eq. \ref{eq:scmf_step4}. If $i_n$ belongs to the second type of sites, then:
    
    \begin{align}
    & \frac{d}{dt}\left\langle n_{i_{1}}^{\alpha_{1}}\dots n_{i_{n}}^{\alpha_{n}}\right\rangle^\mathrm{oe}= \nonumber \\ 
    & \left\langle n_{i_{1}}^{\alpha_{1}}\dots n_{i_{n}}^{\alpha_{n}}\sum_{j,\alpha}n_{j}^{\alpha} \sum_{s\in\theta_{j}^{\alpha}}m_{js}^{\alpha}\omega_{js}^{\alpha}\left(d_{js}^{\mu}\nabla\bar{\mu}_{\alpha}-\sum_{\sigma}\tilde{n}_{\sigma}\nu_{\sigma}\right)\right\rangle. \label{eq:msite1}
    \end{align}
	The sum over jumps can be separated into two contributions: association jumps (i.e., jumps bringing sites $i_1\dots i_n$ to an associated configuration) and jumps between two dissociated configurations. For the latter, $n_c$-body effective interactions are neglected for each dissociated configuration ($\nu_{\sigma}$=0 for both configurations before and after the jump); additionally, detailed balance ensures that the gradient contribution of jumps between dissociated configurations is also zero because $\left\langle \omega_{js}^{\mu}d_{js}^{\mu}\right\rangle=-\left\langle \omega_{sj}^{\mu}d_{sj}^{\mu}\right\rangle$. We are thus left with association jumps only:
    \begin{align}
    & \frac{d}{dt}\left\langle n_{i_{1}}^{\alpha_{1}}\dots n_{i_{n-1}}^{\alpha_{n-1}}\right\rangle^{oe}=\nonumber\\
    & \sum_{i_{n}\not\in c}\left\langle n_{i_{1}}^{\alpha_{1}}\dots n_{i_n}^{\alpha_n}\sum_{j,\alpha}n_{j}^{\alpha}\sum_{s\in\theta_{j}^{\alpha}}m_{js}^{\alpha}\omega_{js}^{\alpha}\tilde{n}_{\sigma}\left(\tilde{n}_{\sigma}d_{js}^{\mu}\nabla\bar{\mu}_{\alpha}-\nu_{\sigma}\right)\right\rangle,\label{eq:msite2}
    \end{align}
    where $i_n$ is a site such that the configuration is dissociated when species $\alpha_n$ is located at site $i_n$ ($i_n\not\in c$), and $\sigma$ is an associated configuration of the cluster. Generally, Eq. \ref{eq:msite2} is not zero, and sub-cluster configurations are therefore not stationary. However, the right-hand side of Eq. \ref{eq:msite2} cancels out in crystallographic systems that are invariant under at least one symmetry operation that reverses the CPG direction and maintains site $i_n$. This is because, in the ensemble average, only configurations where atom $\alpha _{n}$ is located on site $i_n$ are included, and reversing the CPG direction entails a negative $d_{js}^{\mu}$ and a negative $\tilde{n}_{\sigma}$.
    Having no symmetry operation that reverses the CPG direction while preserving lattice sites is equivalent to having symmetric contributions to effective interactions \cite{nastar:2014}, or  to having a non-zero average bias vector on site $i_n$ \cite{trinkle:2017}.
	Most common crystallographic systems are invariant under such symmetry operation (in particular Bravais lattices), and Eq. \ref{eq:msite2} amounts to zero with $n_c$-body effective interactions only. But for some systems (e.g., sheared diamond with a CPG  perpendicular to the shearing plane), sub-cluster effective interactions must be explicitly considered. The details of the calculation of the site interaction contributions are provided in \ref{site_anx}. The derivation amounts to a new expression of transport coefficients, and Eq. \ref{eq:matrixnotation} becomes:
    \begin{equation}
\mathbf{L}_{d}^{\mathrm{eq}}\left(c\right)=\frac{\boldsymbol{\Lambda}_{d}^{0}-\left[\boldsymbol{\Lambda}_{d}\mathbf{\tilde{\mathbf{T}}^{-1}}\boldsymbol{\Lambda}_{\mu}^{t}+\boldsymbol{\gamma}_{d}\boldsymbol{\tau}^{-1}\boldsymbol{\gamma}_{\mu}^{t}\right]}{V},\label{eq:msite3}
\end{equation}
where the $\boldsymbol{\gamma}_{d}\boldsymbol{\tau}^{-1}\boldsymbol{\gamma}_{\mu}^{t}$ term is the correction due to site interactions. Following the notations given in \ref{site_anx}, the correction term is computed using:

\begin{equation}
	\boldsymbol{\lambda}_{d,\alpha\delta}=\left\langle \sum_{\sigma} n_{\sigma} \sum_{j,\alpha} n_{j}^{\alpha} \sum_{s\in \theta_{j}^{\alpha}}m_{is}^{\alpha} \omega_{js}^{\alpha} \tilde{n}_{\not \in}N_{\sigma}d_{sj}^{d}\mathbf{S}_{\mathbf{d},\sigma_{\alpha js}\delta}\right\rangle, \label{eq:msite4}
\end{equation}

\begin{equation}
	\mathbf{D}_{\sigma_0\delta}=\left\langle n_{\sigma_0} \sum_{j,\alpha} n_{j}^{\alpha} \sum_{s\in \theta_{j}^{\alpha}}m_{is}^{\alpha} \omega_{js}^{\alpha} \tilde{n}_{\not \in}N_{\sigma}\mathbf{S}_{\mathbf{d},\sigma_{\alpha js}\delta}\right\rangle, \label{eq:msite5}
\end{equation}

\begin{align}
	& \left(\mathbf{S}_{\mathbf{d}}^{t}\mathbf{\tilde{T}_{\boldsymbol{\delta}}S_{d}}\right)_{\delta_1\delta_2}= \nonumber \\
    & \left\langle \sum_{\sigma} n_{\sigma} \sum_{j,\alpha} n_{j}^{\alpha} \sum_{s\in \theta_{j}^{\alpha}}m_{is}^{\alpha} \omega_{js}^{\alpha} \tilde{n}_{\not \in}N_{\sigma} \mathbf{S}_{\mathbf{d},\sigma_{\alpha js}\delta_1} \mathbf{S}_{\mathbf{d},\sigma_{\alpha js}\delta_2}\right\rangle, \label{eq:msite6}
\end{align}
where $\tilde{n}_{\not \in}=\prod_{\sigma}(1-|\tilde{n}_{\sigma}|)$ and $\mathbf{S}_{\mathbf{d},\sigma_{\alpha js}\delta}$ is the amount (positive or negative integer) of site interactions $\delta$ contained in a dissociated configuration which is similar to configuration $\sigma$, except that one atom of species $\alpha$ has jumped from site $j$ to $s$.

From a physical point of view, site effective interactions can be interpreted as a local relaxation of the macroscopic driving force, in our case the homogeneous CPG. This behavior is similar to the internal relaxation of strain in a non-Bravais mono-crystal undergoing uniform strain (see Ref. \cite{Nastar95} and references therein).

The same reasoning can be applied successively to sub-cluster interactions that are larger than site interactions. For instance, for a cluster of three components in a system where sub-cluster interactions are needed, using site interactions and three-body effective interactions does not guarantee that pair occupation averages are stationary. Hence, two-body interactions should in principle be added, and an equation similar to Eq. \ref{eq:msite2} should be considered. However, in the current version of KineCluE the sub-cluster interaction correction is limited to site interactions only.

    \section{Software implementation \label{sec:techno}}
	
	KineCluE is a set of Python scripts aimed at the computation of the transport coefficients of a cluster, typically consisting of PDs and/or impurities embedded in an infinite lattice. In this respect, it generalizes previous codes dedicated to the application of the SCMF method to specific diffusion mechanisms \cite{PhysRevB.88.134201, messina:2015, schuler:2016}. It runs in a Python 3.6 environment, with no need for high-performance computational tools. It consists mainly of three files: a module containing the definitions of all classes and functions, and two scripts to perform the calculation, first in symbolic and then in numeric form. The former computes the flux equations (Eq. \ref{eq:scmf_step2}) and produces the list of configurations and jump frequencies that need to be considered within the specified interaction radius, whereas the latter performs the actual numerical calculation after the user has provided the relevant binding and saddle-point energies. 
	
	The code is highly versatile in terms of crystal structures, defects, cluster size, and jump mechanisms. It can treat diffusion of interstitial solutes, vacancies, dumbbells, as well as more complex, possibly multi-component defects. Many PD-mediated diffusion mechanisms for substitutional impurities can be modeled, from the traditional to the more complex ones, e.g., the recently theorized half-vacancy mechanism characterizing the diffusion of oversized solute atoms \cite{bocquet:2017} (see Sec. \ref{sec:osa}) . It works for any crystal structure that can be described by some periodic vectors and a set of basis atoms, and in principle for clusters of any size and composition, although there are some practical limitations due to computational time increase with cluster size (see Sec. \ref{sec:computetime}). 
	
	In addition to the tensor of transport coefficients in three diffusion directions, the code computes the cluster partition function, as well as drag ratios, correlation factors, and other relevant cluster properties such as mobilities and dissociation probabilities, quantities that are required for higher scale models such as cluster dynamics and object kinetic Monte Carlo.
	
	\subsection{Overview}
	
\begin{figure}
		\centering{}\includegraphics[width=1\columnwidth]{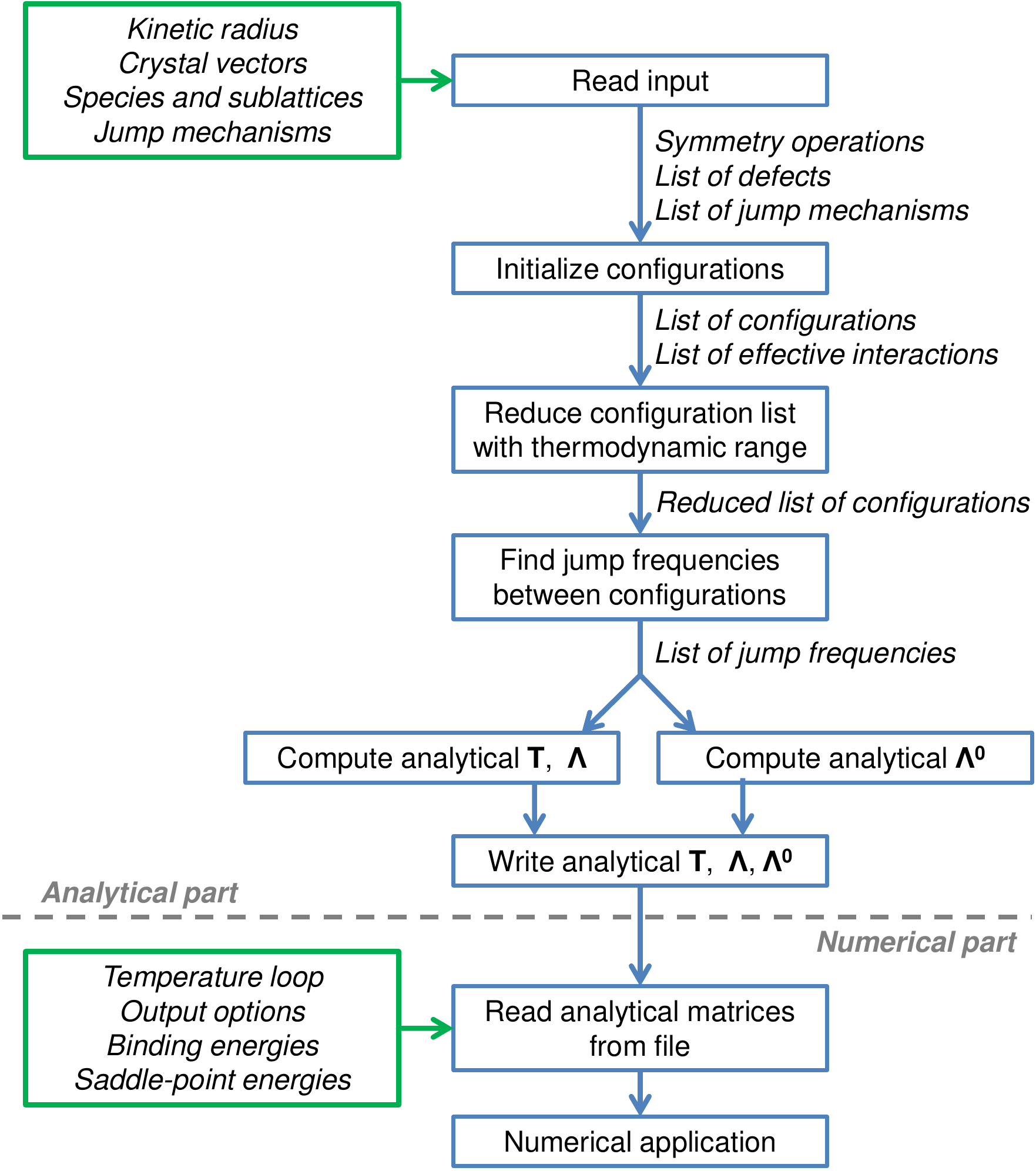}\caption{\label{fig:overview_kineclue}Summary of the main steps of the KineCluE
			code. Green boxes mark user inputs, and blue boxes some
			computation made by the code. The outputs of this computation are
			listed below each box. Note that the code consists of two
			parts: the analytical calculation, and the numerical application to compute the cluster
			transport coefficients.}
	\end{figure}

	The logical diagram of the code is depicted in Fig. \ref{fig:overview_kineclue}. We present here the main features relating the technical implementation with the SCMF theory exposed in the previous section. For more details and practical instructions, we refer the reader to the code documentation.
	
	\paragraph*{User input}
	The main ingredients for the symbolic calculation are: the crystal periodicity vectors and basis atoms, the cluster components and their sublattices (also called "defects"), and the jump mechanisms. Each of these is represented in the code by an instance of the corresponding class (\textit{Crystal}, \textit{Defect}, \textit{Species}, \textit{JumpMech}). In addition, the SCMF method requires the definition of a kinetic radius, i.e. the cutoff distance for effective interactions. To a longer kinetic radius corresponds a higher amount of effective interactions, and hence a larger system of equations; this ensures a better accuracy because more kinetic trajectories are included. Optionally, a second (smaller) cutoff range can be set for thermodynamic interactions, with the aim of speeding up the calculations and reducing the amount of binding and saddle-point energies to compute. This is usually justified by the fact that binding energies between cluster constituents fade quickly with distance. The thermodynamic radius parameter should be set to the minimum value that allows for a correct description of thermodynamic interactions. The optimal choice of the kinetic radius  parameter is discussed in Sec. \ref{sec:convergence}.
	
	\paragraph*{Symmetry operations}
	KineCluE uses crystal symmetries to minimize the amount of effective interactions. At the beginning of the symbolic calculation, the symmetry operations conserving the crystal periodicity are computed using an algorithm inspired by the one by Trinkle \cite{trinkle:2017}, and then applied to find all symmetry-equivalent sublattice positions, cluster configurations, and jumps. In this way, the user needs to specify only sublattices and jump mechanisms that are unique with respect to all possible symmetry operations.

\begin{figure}
		\centering{}\includegraphics[width=1\columnwidth]{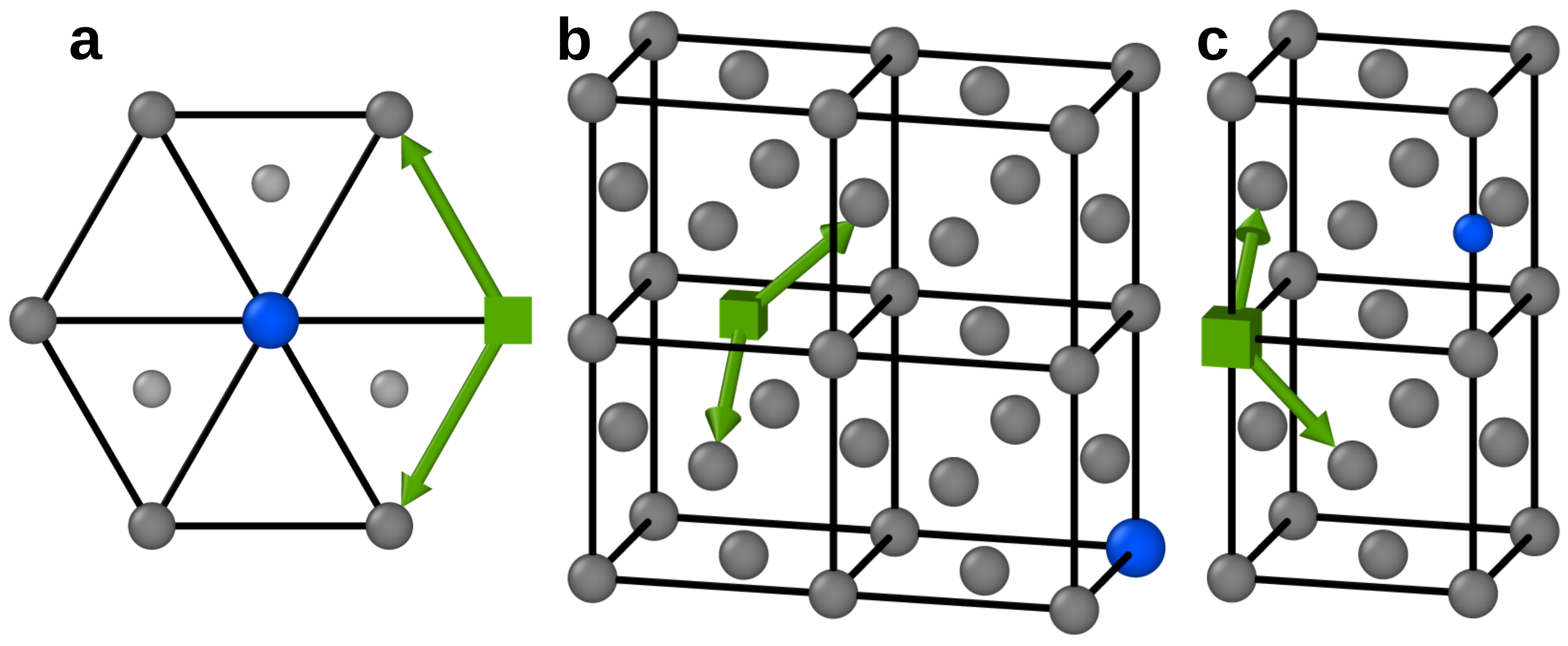}\caption{\label{fig:symmetry_jumps} Examples of vacancy (green square) jumps in the vicinity of a solute (blue sphere) for various lattices: a) hexagonal close-packed, the smaller and lighter spheres representing atoms that are below and above the basal plane; b) face-centered cubic with a substitutional solute; c) face-centered cubic with an interstitial solute. For each system, the two jumps shown with green arrows are not equivalent because of differing saddle-point configurations, even though the initial and final configurations are symmetrically equivalent.}
	\end{figure}

	\paragraph*{Interactions and jump frequencies}
	Symmetry operations are also employed to construct the phase space of cluster configurations (see Sec. \ref{sec:explo}) and classify the configurations in classes of symmetry-equivalent thermodynamic and effective interactions. For each newly found configuration, the symmetry operations are applied to each cluster component to produce all symmetric configurations, which are then assigned the same thermodynamic interaction. Analogously, effective interaction classes are identified by using the subset of symmetry operations that maintain the CPG direction. In addition, symmetries are exploited to determine the list of symmetry-unique jump frequencies. For each allowed jump from an initial to a final configuration, the code applies the symmetry operations on the initial and final positions of the components \textit{simultaneously}. This ensures that jumps with different saddle points but symmetrically identical initial and final positions are properly distinguished, as shown in \cite{agarwal:2017, Nandipati:2016} for hexagonal close-packed structures. While developing the code, the same subtlety was found also in Bravais lattices, an example of which is shown in Fig. \ref{fig:symmetry_jumps}. This emphasizes the need of using crystal symmetry analysis to identify unique transitions.
	
	\paragraph*{Symbolic computation}
	Once the lists of interactions and jump frequencies are finalized, the code proceeds to the symbolic calculation of $\mathbf{T}$ (Eq. \ref{eq:T}), $\boldsymbol{\Lambda}$ (Eq. \ref{eq:lambda}), and $\boldsymbol{\Lambda}^0$ (Eq. \ref{eq:lambda0}) ($\mathbf{M}$ is directly obtained from $\boldsymbol{\Lambda}$, as explained in Sec. \ref{sec:Mlambda}). The $\mathbf{T}$ matrix is built line by line by looping on the list of symmetry-unique effective interactions: for each configuration containing this effective interaction, the code explores all configurations that can be reached with a valid jump, recognizes the effective interaction corresponding to each final configuration, and assigns the transition to the corresponding jump frequencies $\omega_{ij}$. The initial configuration brings in a negative contribution ($-\omega_{ij}$), and the final configuration a positive one ($+\omega_{ij}$) to the matching column of the $\mathbf{T}$ matrix. In the case of homogeneous driving forces, kinetic correlations only depend on the anti-symmetric part of the effective interactions \cite{nastar:2014}. Hence, the contribution is swapped in sign if the symmetry operation leading to this particular instance of the effective interaction class reverses the CPG direction. Concurrently, each column of the $\boldsymbol{\Lambda}$ matrix is obtained by averaging each species displacement (projected on the CPG direction) among all valid jumps from the sample configuration of each effective interaction. The uncorrelated term $\boldsymbol{\Lambda}^0$ is instead obtained by averaging each species net displacement (with no projections) over all possible thermodynamic states of the system. Finally, the code computes the partition function $Z = \sum_{t} g_{t}\exp \left(E^t_\mathrm{b}/k_\mathrm{B}T \right)$, where $t$ marks the thermodynamic interaction, $E^t_\mathrm{b}$ the binding energy (positive means attraction) and $g_t$ the geometric multiplicity, i.e., the number of symmetry equivalents (which fully accounts for the cluster configuration entropy). At this point, the symbolic calculation is complete and the results are stored in a file to be loaded back by the numerical code. 
	
	\paragraph*{Strained systems}
	The symmetry-based approach allows for a straight-forward treatment of strained systems. The user can define a strain tensor to deform the crystal and reduce its symmetry. The code automatically finds the subset of symmetry operations that remain valid, and evaluates if the broken symmetry gives rise to new symmetry-unique sublattices or jump mechanisms that were equivalent in the unstrained system, updating the list of effective interactions and jump frequencies accordingly. The calculation of the symbolic expressions then proceeds in the same way. Strains usually generate anisotropic diffusion terms \cite{PhysRevB.88.134108}, which appear in the code as non-null terms along the directions perpendicular to the CPG one. Note that single-component effective interactions are introduced to apply the correction in Eq. \ref{eq:msite3} for some non-Bravais lattices.
	
	\paragraph*{Numerical evaluation}
For the numerical calculations, the user provides a range of temperature and (optionally) strain values, as well as the binding and saddle point energies for each of the symmetry-unique configurations and jump frequencies found in the symbolic code. The latter are listed in separate files, so that they can be inspected in an atomic visualization software, and computed with the usual methods (density functional theory, interaction models, interatomic potentials, \textit{etc.}). For strain calculations, the user needs to provide as well values of the elastic dipoles of each equilibrium and saddle-point configuration; the energy variation due to elastic energy is computed automatically within linear elasticity theory \cite{Varvenne:2013}. The calculation at this point simply consists in solving numerically the system of equations in Eq. \ref{eq:scmf_step4}, and combining the results with matrices $\boldsymbol{\Lambda}$ and $\boldsymbol{\Lambda}^0$ as in Eq. \ref{eq:matrixnotation}. Additional options allow the user to perform parametric studies as functions, for instance, of the kinetic radius (see Sec. \ref{sec:convergence}), or of specific jump frequencies, as well as sensitivity studies to identify the jump frequencies that have the largest impact on the cluster transport coefficients (see Sec. \ref{sec:sensitivity}).

	After computing the coefficients for each cluster (one calculation each), the total transport coefficients in Eq. \ref{eq:clustercoeff} must be computed outside the code as a post-processing step, depending on the cluster concentrations $[c]$ that can be obtained either in equilibrium or non-equilibrium conditions, for instance by linking with a cluster-dynamics model. 
	
	\paragraph*{Computational load} From a computational perspective, the code is light and can easily run on personal laptops. Distributed under an LGPL license, it can be downloaded freely together with the user documentation and a set of input file examples \cite{KineCluE_rep}. The computational time can be as short as one second for two-component clusters, and increases with cluster size, kinetic radius, and the amount of jump mechanisms. Increasing the kinetic radius or the number of components involves in both cases a wider configuration space and a higher amount of kinetic equations, the construction of which is the bottleneck of the symbolic calculation. See Sec. \ref{sec:computetime} for a more detailed evaluation of the computational time with cluster size and kinetic radius.
	
	\paragraph*{Current limitations} Among the limitations of the current version of KineCluE, it is worth mentioning the following ones.
	\begin{enumerate}
	\item Calculations proceed in the canonical ensemble (fixed number of atoms for each species). This entails that users must keep track of the number of matrix atoms involved in the system, even though they are not declared explicitly. One example is an interstitial solute moving to a substitutional site where an implicit matrix atom was located. As in a Monte Carlo simulation in the canonical ensemble, this matrix atom cannot disappear, and must become for instance a self-interstitial atom. 
	\item  In addition, also the number of defects and solutes in the cluster is fixed. For instance, recombinations between self interstitials and vacancies cannot be modeled, because such a reaction would lead from a two-component cluster (the two PDs) to no defects. This limitation can sometimes be overcome by allowing two components to be located on the same site (see the code documentation, and the kick-out mechanism described in Sec. \ref{sec:diamond}).
	\item At the moment, dumbbell calculations are limited to clusters containing one solute only, because the possible dumbbell compositions currently implemented are limited to matrix- (A-A) and mixed dumbbells (A-B). 
	\item No jump mechanisms involving transmutation reactions (i.e., change of defect type or solute species from initial to final position) are allowed. 
    \item Sub-clusters interactions are limited to site interactions only, whereas they should rigorously extend up to $(n_c -1)$-body interactions. This limitation must be kept in mind when dealing with clusters of 3 components or more in certain non-Bravais lattices. However, we do not expect this approximation to cause a large difference in the value of cluster transport coefficients.
	\end{enumerate}
	
	\subsection{Configuration space exploration algorithm \label{sec:explo}}
	We present in more details our -- to the best of our knowledge -- original algorithm to explore and build the cluster configuration space. Such an algorithm can be useful outside the scopes of KineCluE, for instance to sample the configuration space and perform exact thermodynamic averages, in replacement of traditional rigid-lattice Monte Carlo simulations. From a starting configuration (that can be either provided by the user or automatically generated) the algorithm applies the user-defined jump mechanisms to move from one configuration to the next one. In this sense, the configuration space can be seen as a graph where the nodes (the configurations) are connected with each other by a jump. Symmetry operations are used at each newly found configuration to ensure that symmetry-equivalent configurations are visited only once.
	
	\begin{figure}
		\begin{centering}
			\includegraphics[width=1\columnwidth]{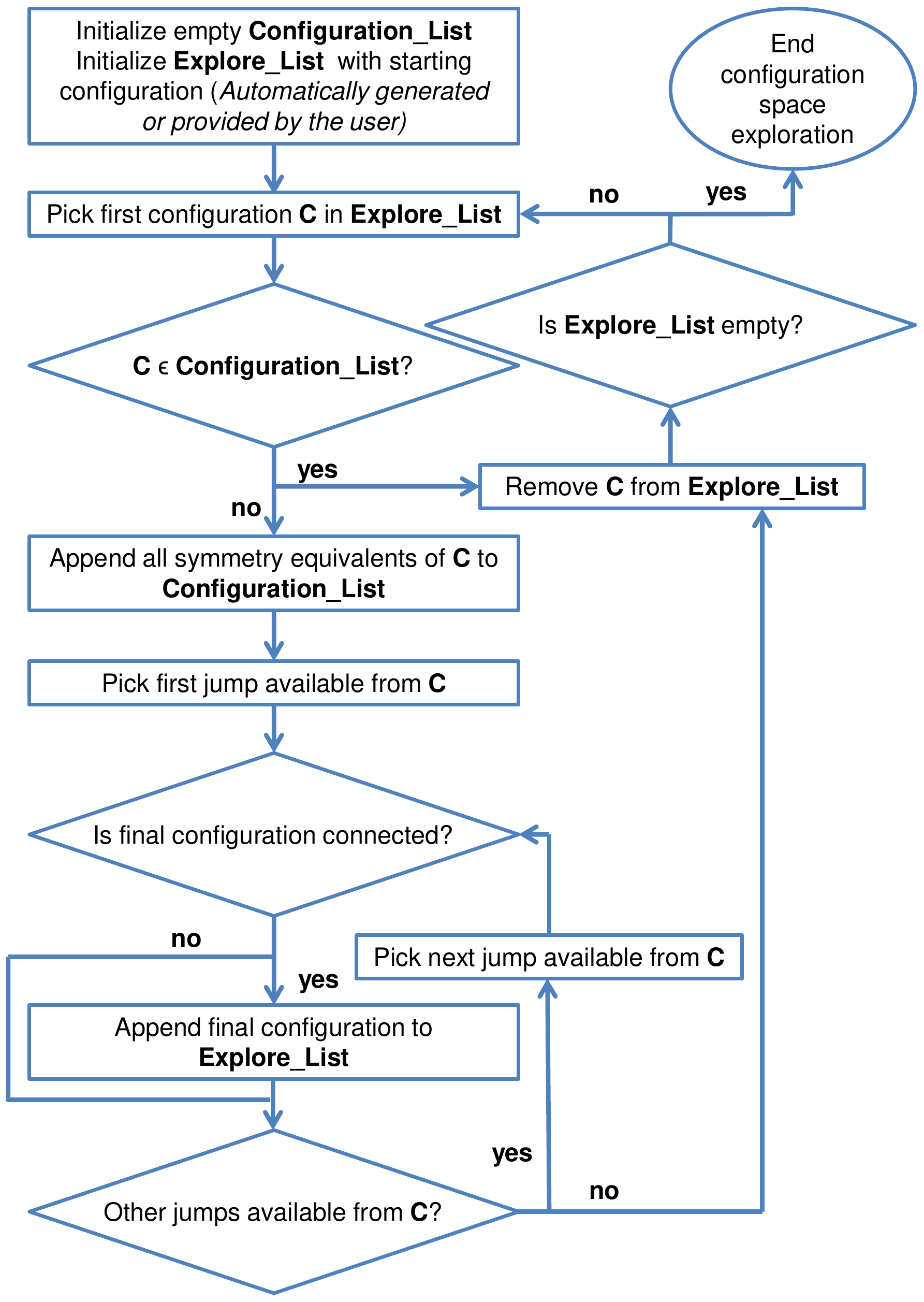}\caption{\label{fig:algo_explo}Algorithm for the exploration of the configuration space, using symmetry operations to minimize the number of configurations to explore.}
			
			\par\end{centering}
		
	\end{figure}
	
	Figure \ref{fig:algo_explo} summarizes the main steps of the algorithm. Configurations to analyze are progressively appended to an exploration list, which initially includes the starting configuration only, and all found configurations are stored in the final configuration list. The exploration stops when the exploration list is empty. For each new configuration $C$ in the exploration list, the code first generates all symmetry-equivalent configurations and appends them to the configuration list, then considers all jumps applicable from $C$. Each final configuration reached by the latter jumps is added to the exploration list, unless it is found to be disconnected, i.e., if any component of the cluster is found at a distance larger than the defined kinetic radius from any other cluster component. Therefore, if the starting configuration is the most compact one, the algorithm builds the configuration space by moving progressively to configurations where the cluster constituents are further and further apart, until one or more of them are beyond the kinetic radius. However, the starting configuration does not need to be the most compact one: if defects, species permissions, and jump mechanisms have been well defined, the resulting configuration space should be well connected and independent from the chosen initial configuration. This is also ensured by the fact that reverse jumps are added automatically by the code if they are not found by symmetry operations. 
	
	Finally, it should be noticed that a correct definition of the jump mechanisms is crucial for the algorithm to visit all configurations existing within the defined kinetic radius, and that not necessarily all the connected configurations are reachable with a given set of jumps. For instance, in the case of a dumbbell-solute pair with radius equal to the 1NN distance, the so-called tensile configurations (solute in a non-target 1NN position) are not accessible, unless an on-site rotation jump is defined \cite{barbe:2006c}.

	\section{Performance and functionality assessment\label{sec:perf}}

    In this section we assess the technical performance and current limitations of the code, discuss the convergence of transport coefficients with respect to the kinetic radius, and explain how to perform sensitivity studies to identify the most important jump frequencies of the system. 
	
	\subsection{Assessment of computation time and memory}
	\label{sec:computetime}
	
\begin{figure}
	\begin{centering}
		\includegraphics[width=1\columnwidth]{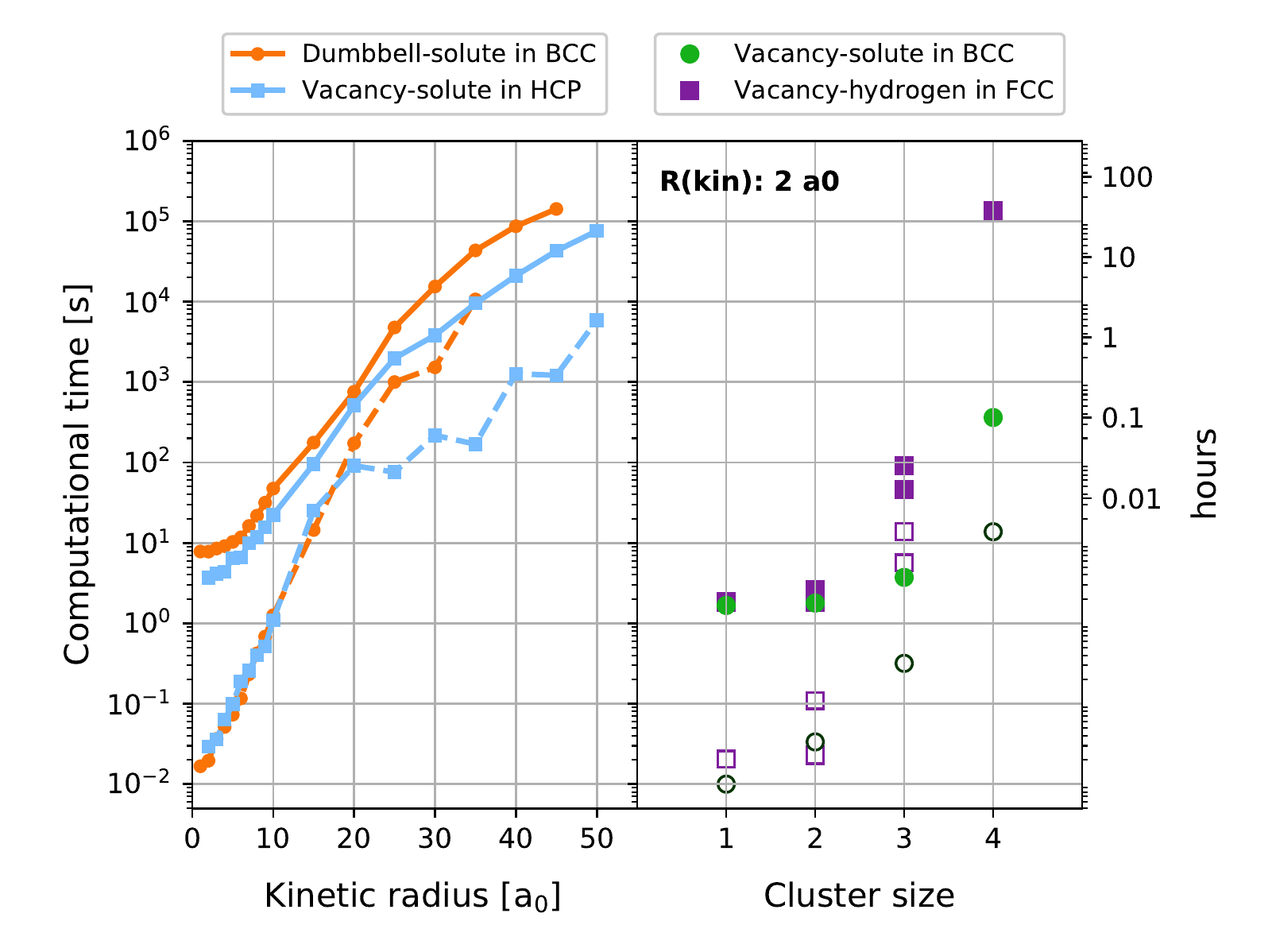}
		\caption{\label{fig:comp_time}Measured computational times as functions of kinetic radius (left, expressed in units of lattice parameter $a_0$) and cluster size (right) on an Intel\textsuperscript{\textregistered} Xeon\textsuperscript{\textregistered} CPU E5-2680 v4 (2.40 GHz). The measurements of numerical calculations, marked with dashed lines and empty markers, refer to the average time per computed temperature data point.}
		\par\end{centering}
\end{figure}

In the analytical code, the bottleneck of the calculation is in the construction of the configuration space and of the analytical system of equations. Therefore, the parameters most affecting the computational performance are the kinetic radius and the cluster size, because they are both proportional to the amount of configurations and of kinetic interactions. Figure \ref{fig:comp_time} shows measurements of the computational time as functions of these two quantities, in the following sample cases: vacancy-solute and dumbbell-solute pairs (kinetic radius, left panel); clusters of vacancies and substitutional or interstitial solutes (cluster size, right panel). The tests were run on a workstation equipped with an Intel\textsuperscript{\textregistered} Xeon\textsuperscript{\textregistered} CPU E5-2680 v4 processor (2.40 GHz). The timing of the numerical code was averaged on the amount of temperature-strain data points.

For both parameters, the computation time increases quickly on the logarithmic scale. Concerning the kinetic radius, this is of no issue because results are usually already well converged after a few lattice parameters, as is showed in the next section. It is thus unneeded to go to very large radii, and in the vast majority of cases it is  possible to run calculations in a matter of seconds or  minutes, or in any case in no longer than an hour. 

On the contrary, computation time represents at the moment a technical limitation for the cluster size that can be handled in KineCluE. It will be necessary, in order to extend its capabilities to larger clusters, to implement parallelization strategies on the construction of the system of equations and of the configuration space. Furthermore, another limitation is given by memory requirements: in Fig. \ref{fig:comp_time}, the tests could not be performed beyond 45-50 $a_0$ and 4-5 components because the RAM memory of the workstation was completely filled up. The main limiting factors are also in this case the amount of configurations and effective interactions: for instance, a computation with 1.08 million configurations and 136 thousands interactions required about 2.4 GB of RAM. This issue will be addressed in future versions of the code. 

In conclusion, the current version of KineCluE is limited to clusters of size 4 or 5, depending on the chosen kinetic radius, whereas for pairs it is possible to obtain very well converged results in reasonably short times on standard personal workstations. 
	
	\subsection{Convergence of pair transport coefficients} \label{sec:convergence}
	
\begin{figure}
		\centering{}\includegraphics[width=1\columnwidth]{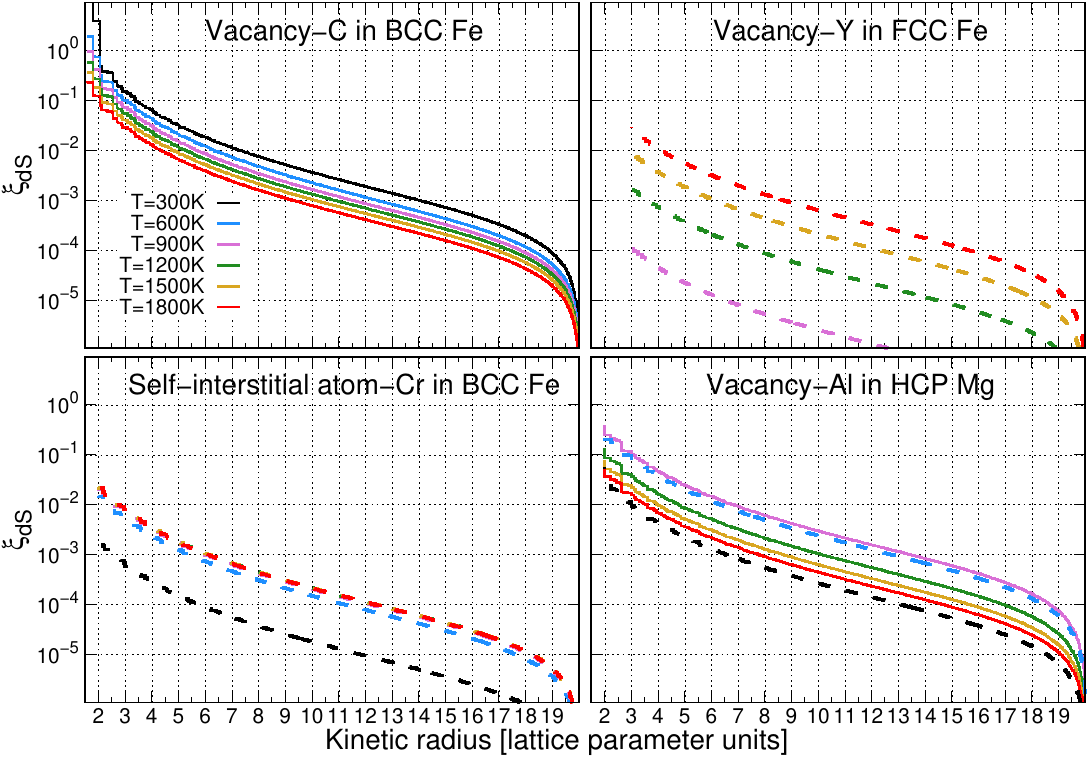}\caption{\label{fig:convergence} Convergence of cluster transport coefficients as a function of the kinetic radius for various crystallographic systems and jump mechanisms. $\xi_{dS}$ is the absolute value of the relative error on the $Z_{dS}L_{dS}(dS)$ product, the reference value being taken at a kinetic radius of 20 lattice parameters. Solid (resp. dashed) lines represent positive (resp. negative) values of the relative error. Note that in the bottom-left plot, all curves from $T=900$ K upwards are superimposed.}
	\end{figure}

	The larger the kinetic radius, the more kinetic trajectories are included in the calculation. At infinite kinetic radius, all possible trajectories are included so that the formalism adopted in KineCluE should theoretically converge towards the exact value for an isolated cluster in an infinite medium. This paragraph shows that cluster transport coefficients converged within practical errors are obtained for kinetic radii no larger than a few lattice parameters. The investigation is summarized in Fig. \ref{fig:convergence}, and was performed for various crystallographic systems, jump mechanisms, and temperatures, each informed by previously published \textit{ab initio} calculations \cite{agarwal:2017,Barouh2015,bocquet:2017,messina:2015}, so the convergence speed varies from one system to another.
	
	In a dilute system containing only monomers and pairs, all the off-diagonal contributions to the Onsager matrix come from interactions within the defect-solute ($dS$) pair cluster. Hence, we know that the product between the off-diagonal cluster transport coefficient $L_{dS}(dS)$ and the pair partition function $Z_{dS}$ converges towards a well-defined physical quantity. Therefore, we choose the absolute value of the relative error on the $Z_{dS}L_{dS}(dS)$ product to evaluate convergence and to choose the most appropriate kinetic radius, which should: 1) be sufficiently high so that all important kinetic trajectories are taken into account (converged quantity); 2) be sufficiently low so that all clusters in the solid solution remain isolated from each other (dilute hypothesis, see Sec. \ref{sec:cluster}). Once the kinetic radius is set for the pair cluster, it should be kept identical for all clusters in the system, otherwise illogical configurations would arise, for instance a three-body configuration where all components interact as pair clusters but the configuration is not considered as a three-body cluster (or the opposite).
	
	Figure \ref{fig:convergence} shows the absolute value of the relative error on $Z_{dS}L_{dS}(dS)$ with respect to its value at $R_\mathrm{kin}=20$ a\textsubscript{0} taken as reference. First of all, this contribution is indeed converging monotonically with increasing kinetic radius. Second, the sign of the relative error (shown as solid and dashed lines) varies from one example to another, but it is correlated with the actual sign of the off-diagonal transport coefficients. In the end, choosing a lower kinetic radius always leads to underestimated values of the absolute value of the off-diagonal transport coefficients, which is logical since these coefficients mainly contain contributions from correlated trajectories. The larger the kinetic radius, the more of these correlated trajectories are added to the calculation, even though their thermodynamic weight decreases with increasing distance. This is consistent with findings from variational approaches to transport coefficients \cite{spohn:1991,arita:2018}. Another interesting point is the convergence evolution with temperature. We found that increasing the temperature always leads to a decrease in the value of the off-diagonal transport coefficients, such that the relative error increases if these coefficients are positive, while it decreases if they are negative. This behavior is observed in Fig. \ref{fig:convergence} by looking at the order of the curves depending on whether they are solid or dashed. Finally, it is noteworthy that all relative errors drop below 10\% at a kinetic range of three lattice parameters. Although this may not be true for other examples, the variety of jump mechanisms and interactions strongly points towards the validity and applicability of our cluster transport coefficients formalism to a wide variety of systems. 
	
	\subsection{Sensitivity study\label{sec:sensitivity}}
	In this section we demonstrate how to use the sensitivity study routine of KineCluE, and how it can be useful to an efficient parameterization of the kinetic model efficiently. As an example, we consider the diffusion of a vacancy-carbon pair ($V$C) in Fe, and rely on previously published \textit{ab initio} data \cite{Barouh2015}. 
	After the analytical code, a set of jump frequencies are identified, and the user needs to supply numerical values for the saddle-point energies of all of them. If saddle-point energies are not provided, KineCluE will use by default the kinetically-resolved activation (KRA) barrier approximation \cite{VanderVen2005a} based on the binding energies of the initial and final state (in our example we assume that all binding energies have been computed accurately):
	\begin{equation}
	E_\mathrm{sp}^{\alpha}=Q^{\alpha}-\frac{E_\mathrm{b}^\mathrm{ini}+E_\mathrm{b}^\mathrm{fin}}{2},
	\label{eq:kra}
	\end{equation}
where $E_\mathrm{sp}^{\alpha}$ is the saddle-point energy for a jump performed by species $\alpha$ between two configurations which are characterized by binding energies $E_\mathrm{b}^\mathrm{ini}$ and $E_\mathrm{b}^\mathrm{fin}$ (positive binding energy means attraction). $Q^{\alpha}$ is an activation energy specific to a jump mechanism of species $\alpha$, usually taken as the migration energy of this species taken as isolated in an infinite bulk system, and using this particular mechanism to diffuse.
	
\begin{figure}
		\centering{}\includegraphics[width=1\columnwidth]{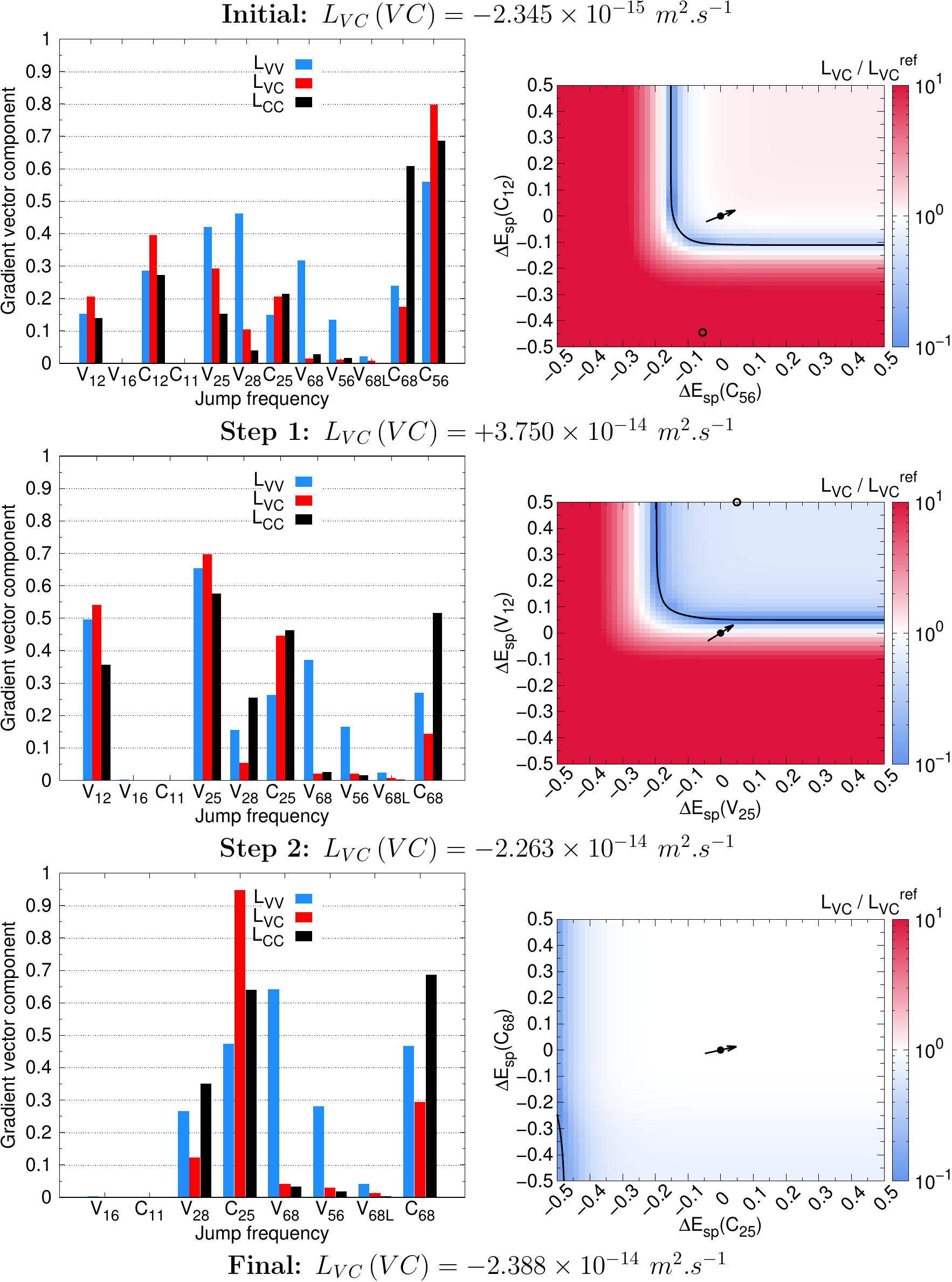}\caption{\label{fig:sensi} Example of sensitivity analysis using \textit{ab initio} data \cite{Barouh2015} for a vacancy ($V$)-carbon (C) pair in Fe at $T=600$ K. Each row corresponds to a step of the sensitivity study, the left-hand plot being the analysis of the most critical jump frequencies (i.e., the ones with the largest components in the normalized gradient vector), and the right-hand plot being the computation of transport coefficients for various saddle-point energy values for the two most critical jump frequencies. In the latter, the axes show saddle-point energy differences with respect to the KRA-computed value, while the color surface shows the absolute value of the ratio between the corresponding $L_{V\mathrm{C}}$ and its reference value before this step of the sensitivity study. The black contour indicates a change in the sign of $L_{V\mathrm{C}}$. The filled circled represents the saddle-point energy values at which the reference $L_{V\mathrm{C}}$ was computed (0,0) and the arrow points towards the direction of highest variation of $L_{V\mathrm{C}}$. Finally, the open circle shows the value of saddle-point energies differences once they were computed with \textit{ab initio} methodds \cite{Barouh2015}. Jump frequencies are labeled $\alpha _{ij}$ for a jump where species $\alpha$ jumps between a configuration where $V$ and C are $i^{th}$ nearest-neighbors and a configuration where they are $j^{th}$ nearest-neighbors.}
	\end{figure}

	It is not possible to tell beforehand how well the KRA approximation performs for a given jump, and a change in saddle-point energies may lead to qualitative changes in transport coefficients. Therefore, we use the KRA value of saddle-point energies as a starting point for a sensitivity analysis. The theory behind this analysis is described in details in \ref{sensi_anx}. It is an iterative process, and each row of Fig. \ref{fig:sensi} corresponds to one step of the process. Each step itself consists of two parts. The first part is the computation of the gradient of each cluster transport coefficients in the jump frequency space, which is simply the partial derivative of said coefficient with respect to each jump frequency. The larger the partial derivative, the larger the change in transport coefficients resulting from a variation in the saddle-point energy of the corresponding jump frequency. Note that dissociation jump frequencies should be removed from this analysis. Indeed, numerous jump frequencies can be assigned to a dissociation frequency, e.g., all jumps between the thermodynamic and kinetic radius. This way, dissociation jump frequencies are made artificially important because of how we regroup jump frequencies together to speed up the calculation. Moreover, they should in principle represent jumps where cluster components do not interact anymore, so that the KRA approximation in Eq. \ref{eq:kra} is expected to hold. The second part is to compute the values of cluster transport coefficients for different values of the most critical jump frequencies, which helps in deciding if it is worth computing these jump frequencies accurately, for instance using \textit{ab initio} methods.
	
	The left-hand side column of Fig. \ref{fig:sensi} shows the normalized gradient at each step, and allows for the identification of the most important jump frequencies. Obviously, some jump frequencies affect to a greater extent some coefficients of the Onsager matrix while not having much effect on other coefficients. To simplify the discussion in this example, we focus on the off-diagonal coefficient at $T=600$ K. The first step of the calculation identifies jump frequencies $C_{56}$ and $C_{12}$ as having the most effect on the $L_{V\mathrm{C}}$ coefficient. The right-hand side column of Fig. \ref{fig:sensi} shows the ratio between the $L_{V\mathrm{C}}$ coefficient for various values of these two jump frequencies and the current $L_{V\mathrm{C}}$ value. We see that variations of 0.2 eV in saddle-point energies can produce a change in the qualitative nature (i.e., sign) of flux coupling between vacancy and carbon. It is thus important to have an accurate estimation of these jump frequencies. In Ref. \cite{Barouh2015} -- using \textit{ab initio} calculations -- the saddle-point energies are found respectively -0.055 eV and -0.445 eV higher than the KRA-predicted value, which leads to a change of sign of the $L_{V\mathrm{C}}$ coefficient as well as to a one-order-of-magnitude increase of its absolute value. Going to the second iteration of the process, jump frequencies $C_{56}$ and $C_{12}$ are set to their \textit{ab initio} values and thus removed from the analysis. Jump frequencies $V_{25}$ and $V_{12}$ are now identified as the most critical ones. Again, the right-hand side plot shows that a small variation of the $V_{12}$ jump frequency value could lead to a change of sign of $L_{V\mathrm{C}}$, so this jump frequency needs to be calculated accurately. On the contrary, jump frequency $V_{25}$ does not seem to have much effect on $L_{V\mathrm{C}}$ unless the saddle-point energy is found about 0.3 eV lower than the current KRA value. In fact, \textit{ab initio} calculations show that the $V_{12}$ jump is not possible (which leads to a change of sign of $L_{V\mathrm{C}}$), and that the $V_{25}$ saddle-point energies is only 0.05 eV higher than the KRA-predicted value. Moving on to the third step and removing jump frequencies $V_{25}$ and $V_{12}$ from the analysis, jump frequencies $C_{25}$ and $C_{68}$ are identified as the most critical ones. But when we look at the right-hand side plot, we see that the impact of these two jump frequencies is rather small, unless the KRA approximation is off by about 0.5 eV for both jump frequencies, which is unlikely. Hence we can consider that we have already computed all the most important jump frequencies, and that the parameterization step can be stopped here. 
	
	On this simple example where the code identified 12 jump frequencies, the sensitivity analysis shows that only 4 of them are needed to get a reasonable estimation of the $L_{V\mathrm{C}}$ coefficient. The value obtained after step 2 ($L_{V\mathrm{C}}=-2.263\times 10^{-14}$ $m^2s^{-1}$) is found to be within a 5\% relative error with respect to the coefficient obtained when all 12 jump frequencies were computed \textit{ab initio} ($L_{V\mathrm{C}}=-2.388\times 10^{-14}$ $m^2s^{-1}$). In larger clusters were hundreds of jump frequencies are identified by the code, this sensitivity analysis can result in important computational savings for the parameterization of the model while having accurate cluster transport coefficient values. 
	
	\section{Benchmark with literature studies\label{sec:tests}}
	
	We present in this section a few validation studies in a broad range of crystal lattices and diffusion mechanisms, including strained systems.
	
	\subsection{Self-diffusion correlation factors in various systems}
	\label{sec:correl}
	
	A first validation consists in computing the self-diffusion correlation factors $f_0$ for several diffusion mechanisms. This is easily done in KineCluE by defining the tracer atom as a foreign species having the same exact energetics as the matrix host atoms. The latter entails null interaction energies, and jump frequencies equal to those in the pure crystal.
	
    In this analysis, a correlation factor is defined as the ratio between a transport coefficient and its uncorrelated part, i.e., its reduction to the $\mathbf{\Lambda}^{0}_{d}$ component in Eq. \ref{eq:matrixnotation}. The uncorrelated part of a transport coefficient is a sum over all possible configurations, and possible jumps from these configurations, of the associated jump frequencies weighted on squared jump distances. In most cases, this definition of the uncorrelated contribution is consistent with the definition given by a random walk. However, when defects visit different non-equivalent configurations, users need to define macro-jumps between equivalent configurations to be able to compute a true random walk \cite{bocquet:1983}. This arbitrary choice of macro-jumps may lead to different $f_0$ values because the sum runs over a reduced number of configurations. For instance, considering dumbbell migration in FCC metals, $f_0$ is found equal to  0.439454 \cite{BENOIST1977} or 0.878908 \cite{bocquet:1983} depending on whether the random walk is defined between jumps or macro-jumps consisting of two successive jumps. Our definition based on $\mathbf{\Lambda}^{0}_{d}$ does not require any definition of macro-jumps, which makes it more general and systematic. Note that this ambiguity only affects the $f_0$ coefficient, not the transport coefficients.
    
	Table \ref{tab:f0_results} shows the computed $f_0$ in comparison with previous analytical or Monte Carlo calculations \cite{manning:1964,compaan:1956,chen:2011,BENOIST1977, ishioka:1978, koiwa:1983, bocquet:1983, bocquet:1991, bocquet:2015, bocquet:2017}, and when available with the Green function (GF) method by Trinkle \cite{trinkle:2017}. As opposed to the latter, the results in KineCluE depend on the chosen kinetic radius $R_\mathrm{kin}$, i.e., on the amount of kinetic trajectories included in the calculation. However, with increasing $R_\mathrm{kin}$ the results converge well to the reference values, and it is not necessary to go to very large radii to obtain an already satisfactory precision. For instance, a calculation with $R_\mathrm{kin} = 6\: a_0$ runs in a few seconds and ensures a precision to the third or fourth decimal digit, which is usually fully satisfactory for the purposes of a diffusion study. It is interesting to mention that, in agreement with the convergence analysis in Section \ref{sec:convergence}, the value of $f_0$ systematically decreases with increasing kinetic radius, as the contribution of correlations becomes larger.   
	
	\begin{table*}\scriptsize
		\caption{Self-diffusion correlation factors for various mechanisms, as computed with KineCluE, in comparison with the Green-function (GF) method \cite{trinkle:2017} and previous calculations \cite{manning:1964,compaan:1956,chen:2011,BENOIST1977, ishioka:1978, koiwa:1983, bocquet:1991, bocquet:2015, bocquet:2017}. The third column refers to the absolute error with respect to the reference value in the last column (or the GF method when available), when the calculation is performed with a smaller kinetic radius (6 a\textsubscript{0}). }
		\renewcommand{\arraystretch}{1.3}
		\centering
		\begin{adjustbox}{max width=\textwidth}
		\begin{tabular}{L{5.5cm}|C{2cm}C{1.2cm}|C{2cm}|C{2cm}|C{3.5cm}}
			\hline
			\textbf{Mechanism}	&	\multicolumn{2}{c|}{\textbf{KineCluE} ($\mathbf{R_\textbf{kin}}$)}	&	\textbf{Error at} $\mathbf{R_\textbf{kin} = 6 \; a_0}$ &	\textbf{GF method}\cite{trinkle:2017}	&	\textbf{other calculations}	\\   
			\hline & & & & &\\ [-6pt]   
			\textbf{Vacancy mechanisms}  & & & & & \\
			BCC                   & 0.72719507  & (50 a$_0$)	& $4.8\cdot 10^{-4}$	& 0.72719414  &  0.727194 \cite{koiwa:1983, manning:1964}  \\
			FCC                   & 0.78145371  & (30 a$_0$)	& $2.7\cdot 10^{-4}$	& 0.78145142  &  0.78145142 \cite{koiwa:1983, manning:1964} \\
			Simple cubic      & 0.65310983  & (60 a$_0$)	& $8.5\cdot 10^{-4}$	& 0.65310884  &  0.653109 \cite{koiwa:1983} \\
			HCP base plane  & 0.78121130  & (30 a$_0$) 	  & $7.2\cdot 10^{-4}$ 	  & 0.78120488  &  0.78120489 \cite{ishioka:1978} \\ 
			HCP axial plane	 & 0.78145784  & (30 a$_0$)		& $7.2\cdot 10^{-4}$	& 0.78145142  &	 0.78145142 \cite{ishioka:1978} \\
			Diamond           & 0.50000082  &  (30 a$_0$)	& $9.8\cdot 10^{-5}$	&	0.50000000  &  0.5 \cite{ishioka:1978, manning:1964} \\  
			& & & & &\\ [-6pt]
			\textbf{Dumbbell mechanisms}  & & & & & \\
			$\langle 110\rangle$ in BCC, 60\degree rotation-translation 		& 0.41264390 & (30 a$_0$)	& $3.1\cdot 10^{-4}$ & - & 0.413010 \cite{bocquet:1991} \\
			$\langle 110\rangle$ in BCC, translation 									  & 0.49432350 & (50 a$_0$)	  &	 $1.2\cdot 10^{-3}$	& - & 0.494371 \cite{bocquet:1991}  \\
			$\langle 100\rangle$ in FCC, 90\degree rotation-translation		    & 0.43945498 & (30 a$_0$)	& $1.0\cdot 10^{-4}$  & - & 0.439454 \cite{BENOIST1977}\\   
			& & & & &\\ [-6pt]
			\textbf{Other mechanisms}  & & & & & \\
			Divacancy mechanism in FCC 							& 0.45809698 & (30 a$_0$)   & $3.0\cdot 10^{-4}$ & - & 0.45809434 \cite{bocquet:2015}  \\
			Oversized solute diffusion in BCC     				 & 0.76161942  & (30 a$_0$)  & $1.8\cdot 10^{-3}$ & - &  0.761603 \cite{bocquet:2017} \\ 
			Oversized solute diffusion in FCC     			   	 & 0.79459770  & (30 a$_0$)  & $8.7\cdot 10^{-3}$ & - &  0.787081 \cite{bocquet:2017} \\
            & & & & &\\ [-6pt]
			\textbf{Kick-out mechanisms in diamond}  & & & & & \\
			Tetrahedral ($t_{1-3}$)     & 	0.9696972				  & (20 a$_0$)  & $8.3 \cdot 10^{-6}$ & - &  0.969733 \cite{compaan:1958}, 0.9701 \cite{chen:2011} \\
            Hexagonal ($h_{1,2}$)    & 		0.851764			  & (20 a$_0$)  & $2.4 \cdot 10^{-5}$ & - &  0.8525 \cite{chen:2011} \\
            Hexagonal ($h_{3,4}$)     & 	0.917991				  & (30 a$_0$)  & $3.5 \cdot 10^{-3}$ & - &  0.9269 \cite{chen:2011} \\
            Hexagonal ($h_{5-8}$)     & 		0.9925246			  & (20 a$_0$)  & $4.0 \cdot 10^{-7}$ & - &  0.9927 \cite{chen:2011} \\
            & & & & &\\ [-6pt]
			\textbf{Multiple interstitial mechanisms in diamond}  & & & & & \\
			Tetrahedral     & 		0.4848494			  & (10 a$_0$)  & $7.2 \cdot 10^{-6}$ & - &  0.4850 \cite{chen:2011} \\
            Hexagonal     & 	0.5641879				  & (10 a$_0$)  & $1.4 \cdot 10^{-5}$ & - &  0.5643 \cite{chen:2011} \\
			& & & & &\\ [-6pt]
			\textbf{Vacancy in two-dimensional lattices}  & & & & & \\
			Square lattice					&  0.46697619  & (100 a$_0$)	&  $8.5\cdot 10^{-3}$	& - &  0.46705 \cite{compaan:1956} 	\\
			Hexagonal lattice			  &  0.56009029  & (100 a$_0$) &  $8.5\cdot 10^{-3}$ & - &  0.56006 \cite{compaan:1956} \\
			Honeycomb structure		 &  0.33336978	 & (100 a$_0$) &  $9.5\cdot 10^{-3}$ & - &  0.33333 \cite{compaan:1956} \\ 
			\hline   
		\end{tabular}
		\end{adjustbox}
		\label{tab:f0_results}
	\end{table*}
	
	We have investigated vacancy-exchange mechanisms in several crystals, including two-dimensional ones, and a few dumbbell mechanisms in BCC and FCC alloys. The case of dumbbell-solute pairs in BCC has been more extensively treated in another work \cite{messina:2015}, where KineCluE has allowed for the extension of the previous SCMF framework \cite{barbe:2006c} from 1NN interactions to arbitrarily long kinetic radii. The correlation factor obtained for the rotation-translation mechanism of $\langle100\rangle$ dumbbells in FCC is in full agreement with Monte Carlo simulations \cite{siegel:1982} and Bocquet's earlier analytical model \cite{BENOIST1977}.  
	
	In addition, we have successfully tested the code on a few more complex mechanisms, for which analytical calculations are available, namely: tracer diffusion in FCC via a di-vacancy mechanism \cite{bocquet:2015}, the diffusion of oversized atoms in BCC and FCC alloys \cite{bocquet:2017}, as well as several kick-out and combined dumbbell-direct interstitial mechanisms in diamond \cite{chen:2011}. The latter two cases are discussed in more detail in Sections \ref{sec:osa} and \ref{sec:diamond}, respectively. 
	
	\subsection{Oversized solute mechanism in BCC and FCC crystals}
	\label{sec:osa}
	
	Thanks to the high degree of flexibility in the jump mechanism definition, KineCluE can handle complex diffusion mechanisms, such as for instance the recently discovered diffusion pattern of oversized solute atoms (OSA) in BCC and FCC alloys \cite{bocquet:2017}. According to this mechanism, OSA diffusion does not occur via a direct exchange with vacancies; instead, when neighboring a vacancy, the oversized atom leaves the substitutional position and relaxes towards the center of the empty lattice space, forming a complex with two half vacancies. In this "split-vacancy" configuration, either one of the two half vacancies can exchange with one of the neighboring matrix atoms. This entails the dissociation of the complex and the return of the solute into a substitutional position (Fig. 1 of the reference work \cite{bocquet:2017}). A net displacement of the solute has occurred if the new substitutional position is different from the one before the complex formation. In FCC structures, it is also possible for the half-vacancy to perform non-dissociative jumps within the 1NN triangle (Fig. 3 of the reference work \cite{bocquet:2017}). Bocquet and coworkers developed an analytical framework to compute the solute correlation factor and diffusion coefficient for this new mechanism, and then considered the case of yttrium in BCC and FCC iron by computing the \textit{ab initio} migration barriers, with solute-vacancy thermodynamic interactions limited to 1NN and 2NN sites. 

\begin{figure}
		\centering{}\includegraphics[width=1\columnwidth]{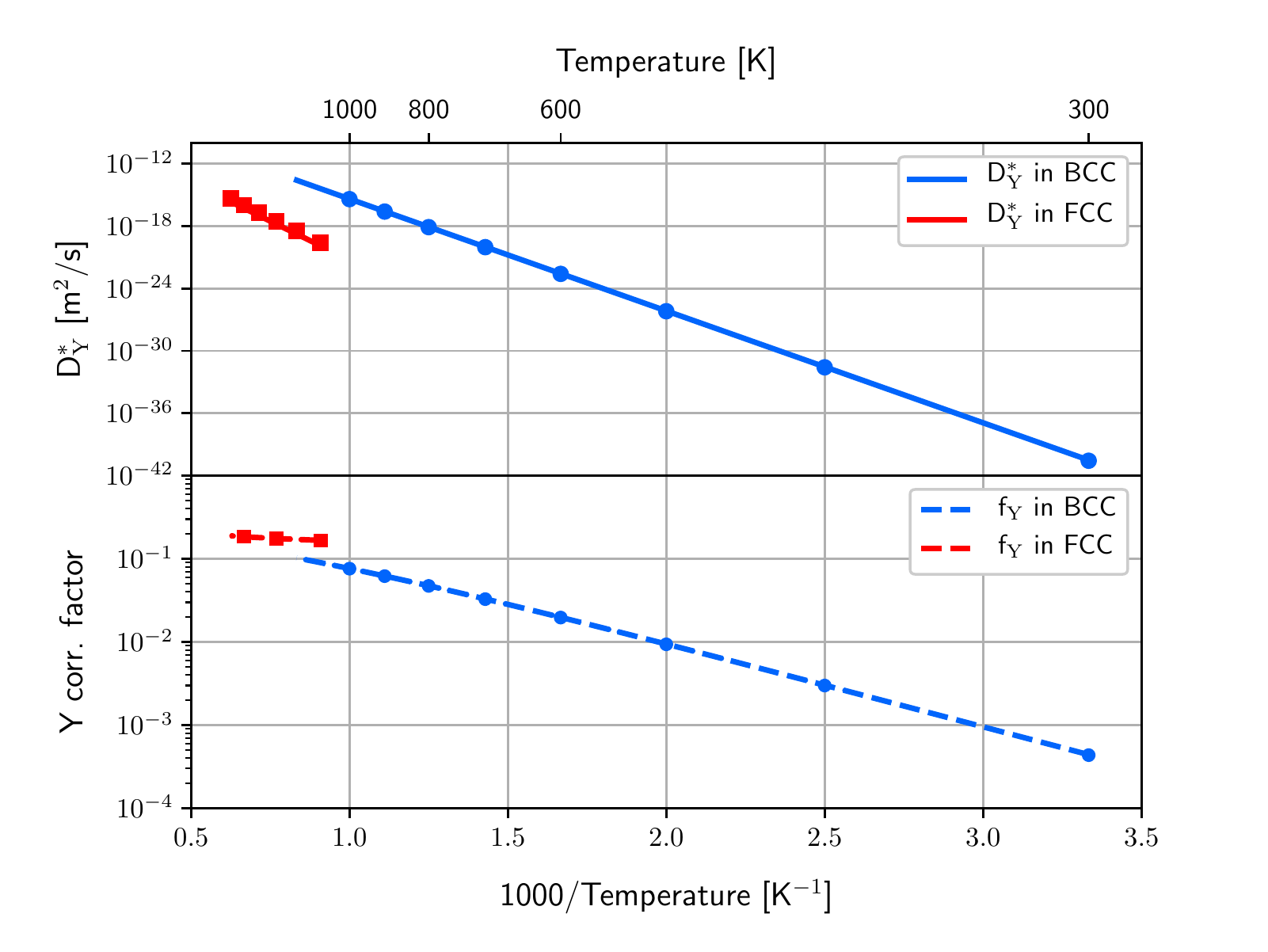}\caption{\label{fig:osa_bocquet}Diffusion coefficient (above) and correlation factor (below) of yttrium in BCC and FCC iron via the oversized solute diffusion mechanism, compared with the analytical calculations (markers) by Bocquet \textit{et al.} \cite{bocquet:2017}. }
	\end{figure}

	With the appropriate definition of configurations and jumps, it is possible to reproduce the same mechanism in KineCluE. The split-vacancy configuration is defined by introducing a second sublattice in the intermediate position between two substitutional atoms, and by setting permissions so that each species (vacancy and solute) is allowed on that sublattice only when sharing the site with the other species. In addition, the usual 1NN substitutional configuration must be forbidden by setting its prefactor to zero in the numerical part of the code. Finally, the solute jump needs to be described with several dissociative jumps, each of which departs from the split-vacancy configuration and brings the vacancy either to a 2NN, 3NN, or 5NN configuration in BCC (2NN, 3NN, or 4NN in FCC). In FCC, it is also necessary to add the 1NN-1NN non-dissociative jump. Sample input files can be found in the code documentation. Figure \ref{fig:osa_bocquet} shows the perfect match between the Y correlation factor and diffusion coefficient obtained with KineCluE and Bocquet's analytical model. A satisfactory agreement is also obtained for the self-diffusion correlation factors shown in Table \ref{tab:f0_results}.
		
	\begin{figure}
		\centering{}\includegraphics[width=1\columnwidth]{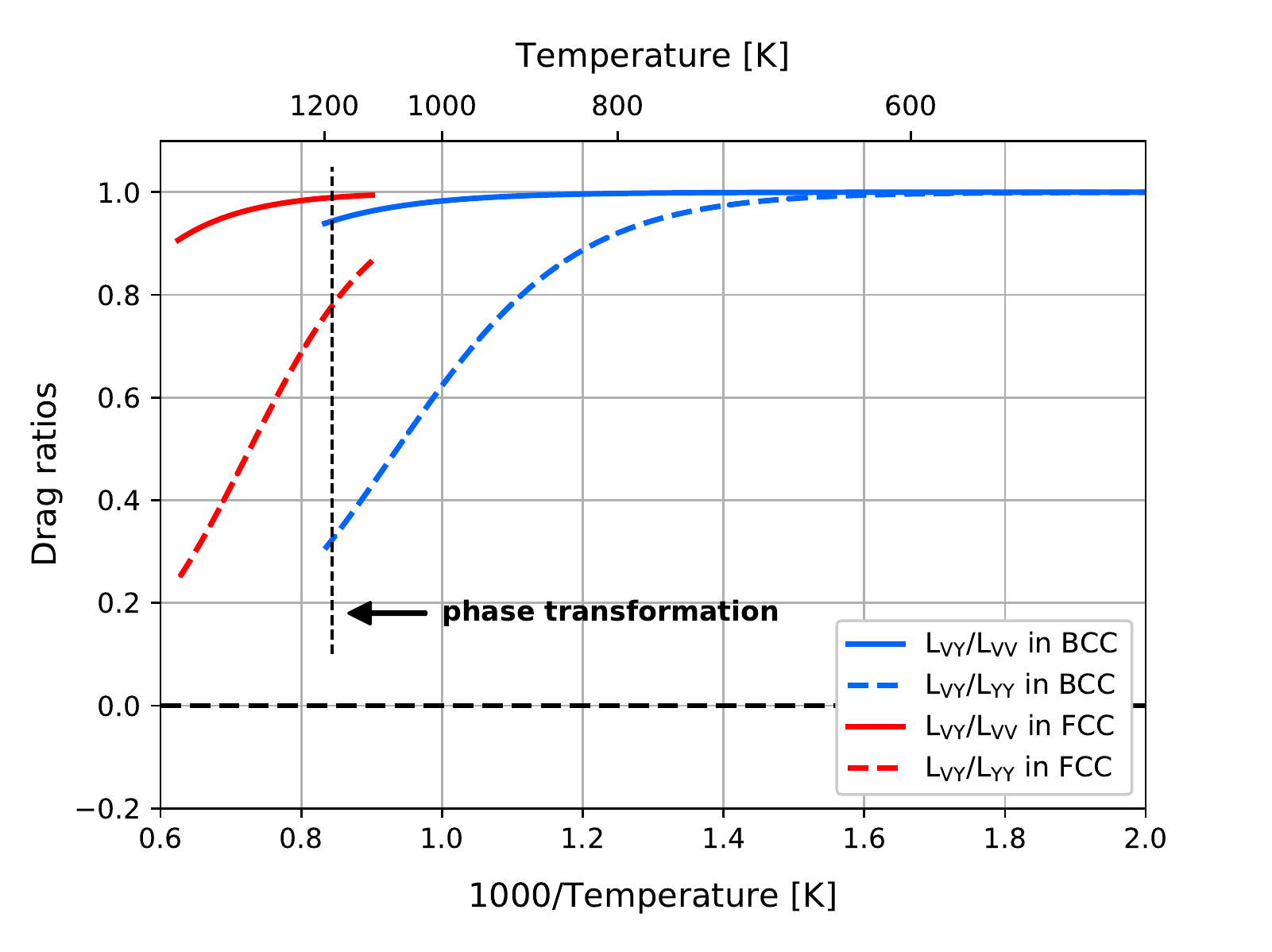}\caption{\label{fig:osa_drag}Variation with temperature of the off diagonal-to-diagonal transport-coefficient ratios, for diffusion of yttrium in BCC and FCC iron via the oversized solute mechanism.}
	\end{figure}
	
		In addition, KineCluE provides a more complete picture of solute-defect correlations thanks to the calculation of the off-diagonal transport coefficient for kinetic radii larger than a few nearest neighbors. We have therefore computed the drag ratios L$_{\mathrm{Y}V}$/L$_\mathrm{YY}$ and L$_{\mathrm{Y}V}$/L$_{VV}$ (Fig. \ref{fig:osa_drag}) with Bocquet's \textit{ab initio} migration barriers, setting the thermodynamic and kinetic radii to 3 a\textsubscript{0} and 4 a\textsubscript{0}, respectively. This leads to 38 distinct jump frequencies in BCC (78 in FCC), but jumps beyond the 2NN shell were set to the isolated vacancy jump frequency. The $L_{VV}$ coefficient is obtained by adding the contribution of the monomers (i.e., single vacancies) to that of the OSA-vacancy pair according to Eq. \ref{eq:clustercoeff}, and computing the pair concentration as $C_{V\mathrm{Y}} = C_{V}C_\mathrm{Y}Z_{V\mathrm{Y}}$ ($C_\mathrm{Y}=1\%$). The partition function $Z_{V\mathrm{Y}}$ is output by KineCluE, while the drag ratios are independent from the vacancy concentration. 
The results show that Y will diffuse by vacancy drag up to very high temperatures and across the phase transformation to FCC. Therefore, not only the diffusion of Y atoms is faster than Fe self diffusion -- as was highlighted by Bocquet -- but it is expected that the highly stable split-vacancy complex leads to strong positive (i.e., same direction) flux coupling between Y solutes and vacancies. Note that the values of the $L_{V\mathrm{Y}}/L_{VV}$ ratios in Fig. \ref{fig:osa_drag} depend in principle on the kinetic radius. Indeed, as the kinetic radius increases, more and more $V$ monomer contributions are included in the $V$-Y pair $L_{VV}$ coefficient. Yet, because there is a strong attractive interaction between $V$ and Y (1.2 eV in BCC and 1.3 eV in FCC), these $V$ monomer contributions are negligible compared with the $V$-Y pair contribution unless the calculation is performed with a very large kinetic radius.
	
	\subsection{Kick-out and multiple interstitial mechanisms in diamond structures}
	\label{sec:diamond}
	
In semi-conductors with a diamond crystallographic structure, kick-out and multiple interstitial diffusion mechanisms having similar activation energies may contribute to atomic transport \cite{bruneval:2012}. KineCluE is able to handle these complex diffusion mechanisms. A frequently invoked mechanism is the one combining hops of a dumbbell and a direct interstitial (described as the "stable-split mechanism" in Ref. \cite{chen:2011} and as a "multiple diffusion mechanism" here).  
The dumbbell is oriented along the $\langle110\rangle$ direction and one of the atoms of the dumbbell is hopping onto a neighboring interstitial site (tetrahedral or hexagonal site).
There is also the so-called kick-out mechanism (or indirect mechanism): an atom sitting on an interstitial site $A$ hops onto a neighboring lattice site $B$ that was occupied by a substitutional atom, kicking the latter out onto a neighboring interstitial site $C$. The interstitial network is either made of tetragonal or hexagonal interstitial sites. For the hexagonal interstitial network, we distinguish three sub-mechanisms depending on the cosine of the angle between vectors $\overrightarrow{AB}$ and $\overrightarrow{BC}$: the first sub-mechanism starting from a given interstitial site $A$ involves two final hexagonal sites $C$ (labeled $(1,2)$ in \cite{chen:2011}) forming an angle with cosine equal to $9/11$, the second sub-mechanism involves two hexagonal sites $C$ labeled $(3,4)$ with a cosine equal to $5/11$, and the third set involves four hexagonal sites $C$ labeled $(5, 6, 7, 8)$ with a cosine equal to $1/11$. The agreement  between the correlation factor $f_0$ obtained from KineCluE (Table \ref{tab:f0_results}) and the values extracted from atomic kinetic Monte Carlo simulations \cite{chen:2011} is very good for both the kick-out and multiple diffusion mechanisms.
Note that in order to compare with data from Ref. \cite{chen:2011}, we have used the measured average cosine value $\cos^\mathrm{tr}\theta$ of the angle between successive jumps of a tracer atom  and deduced the corresponding $f_0$ using the formula: $f_0=1+\cos^\mathrm{tr}\theta$. As shown in Table \ref{tab:f0_results}, with a kinetic radius larger than $10 \; a_0$, KineCluE achieves a better precision because the resulting correlation factors are systematically lower than the corresponding Monte Carlo values. The precision of the Monte Carlo simulation is around $10^{-2}$, accuracy that in KineCluE is already reached with $R_\mathrm{kin}=1.5\;a_0$. Thus, although it was not specified in the reference work, we estimate the size of the Monte Carlo simulation cell in \cite{chen:2011} to be around $2R_\mathrm{kin}=3a_0$. This is based on the idea that kinetic correlations associated with defect trajectories going beyond the size of the Monte Carlo simulation cell are not properly taken into account due to the periodic conditions and the consequent interference of a trajectory with itself.
	
	\subsection{Transport coefficients in strained systems}

KineCluE allows for an essentially seamless application of the SCMF cluster-expansion method to strained systems, thus widely improving the state-of-the-art models for computing fully strain-dependent transport matrices. Strain-dependent diffusion coefficients have been measured by molecular dynamics \cite{Averback2008} and by Monte Carlo simulations \cite{Ziebarth2015}, but it has been shown that measuring the elasto-diffusion tensor with the latter method is tricky \cite{Zebo2016}. Previous analytical models allowing for such kind of calculations were based on:
\begin{itemize}
	\item [-] the SCMF method (before cluster expansion) by Garnier and coworkers, for diffusion of substitutional solutes via a vacancy mechanism in FCC alloys \cite{PhysRevB.88.134108,PhysRevB.90.024306,Garnier:2014_1};
    \item[-] the Green-function method for isolated interstitial solutes \cite{trinkle:2016,trinkle:2017}.
\end{itemize}
KineCluE greatly expands the range of systems that can be treated under strain, taking accurately into account all effects on correlations and energetics; for instance, it can correctly describe the effects of strain on the flux coupling between dumbbells and substitutional solutes, for which no model currently exists. In the analytical part of the code, the symmetry analysis allows for the correct identification of the broken symmetries in the strained crystal, whereas the effect of the elastic energy of the strain field is automatically included in the numerical code within the framework of linear elasticity theory \cite{Varvenne:2013}.

\begin{figure*}
	\centering{}\includegraphics[width=1\textwidth]{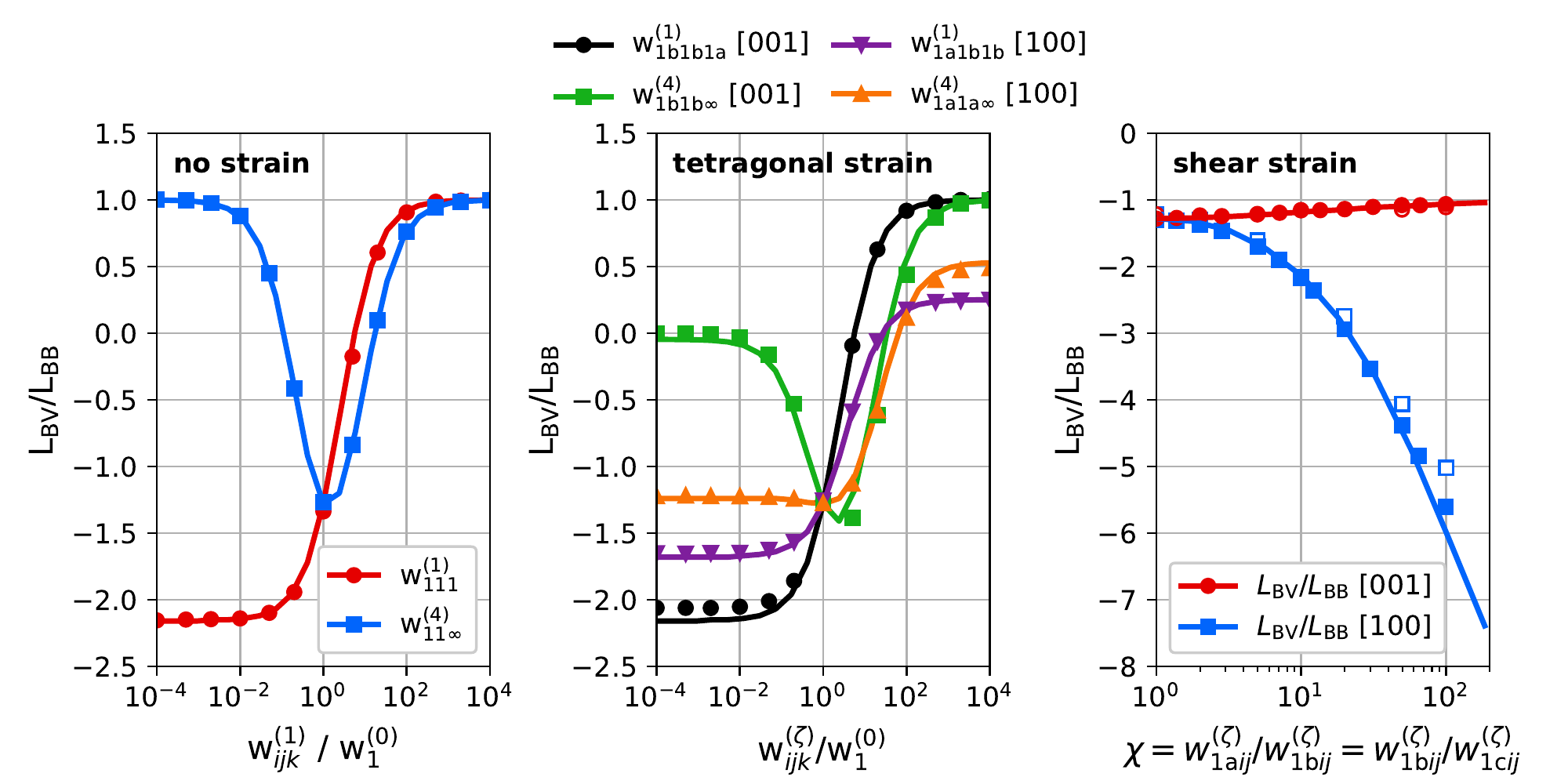}\caption{\label{fig:nisi} Drag ratios for substitutional solute diffusion by vacancies in an FCC model alloy without strain (left), under a tetragonal strain $\varepsilon_{33}$ (middle), and under a shear strain $\varepsilon_{12}$ (right), as functions of different frequency ratios, compared with the SCMF calculations of a previous work \cite{Garnier:2014_1} (full markers) and using the nomenclature therein. Empty markers report the corresponding kinetic Monte Carlo calculations in the same work. Each strain calculation refers to a specific diffusion direction (in the shown cases, directions [100] and [010] are equivalent). }
\end{figure*}

The implementation has been successfully tested on the aforementioned studies. Figure \ref{fig:nisi} shows the drag ratio $L_\mathrm{BV}/L_\mathrm{BB}$ for a vacancy-solute pair in an FCC model alloy, as a function of different jump frequency ratios in the unstrained crystal (left panel), under a tetragonal strain $\varepsilon_{33}$ (middle panel), and under a shear strain $\varepsilon= 2\varepsilon_{12}=2\varepsilon_{21}$ (right panel). The nomenclature is the same as in the reference SCMF-based work \cite{Garnier:2014_1} (Fig. 1 and 2 therein). The diffusion direction (e.g. [001]) is controlled in KineCluE with the CPG direction, and by optionally adding two other non-collinear diffusion directions in case non-isotropic terms arise. The calculations were performed with a larger kinetic radius ($4\:a_0$) than Garnier's model ($\sqrt{2}\:a_0$), which entails larger correlations (larger $|L_\mathrm{BV}|$) and slower solute diffusion (smaller $L_\mathrm{BB}$). Hence, the drag ratios computed with KineCluE are larger in absolute value then the ones computed by Garnier, as shown in Fig. \ref{fig:nisi}. In the example of $L_\mathrm{BV}/L_\mathrm{BB}\;[100]$ under shear strain, atomic kinetic Monte Carlo simulations \cite{Garnier:2014_1} performed in a $6\: a_0 \times 6\: a_0\times 6\: a_0$ cell (corresponding approximately to a kinetic radius of $3\:a_0$) are expected to appear between the results from KineCluE ($R_{kin}=4\:a_0$) and from Garnier ($R_{kin}=\sqrt{2}\:a_0$). However, this is not the case, most likely because of statistical uncertainties, especially as the difference appears at large jump frequency ratios.

\begin{figure}
	\centering{}\includegraphics[width=1\columnwidth]{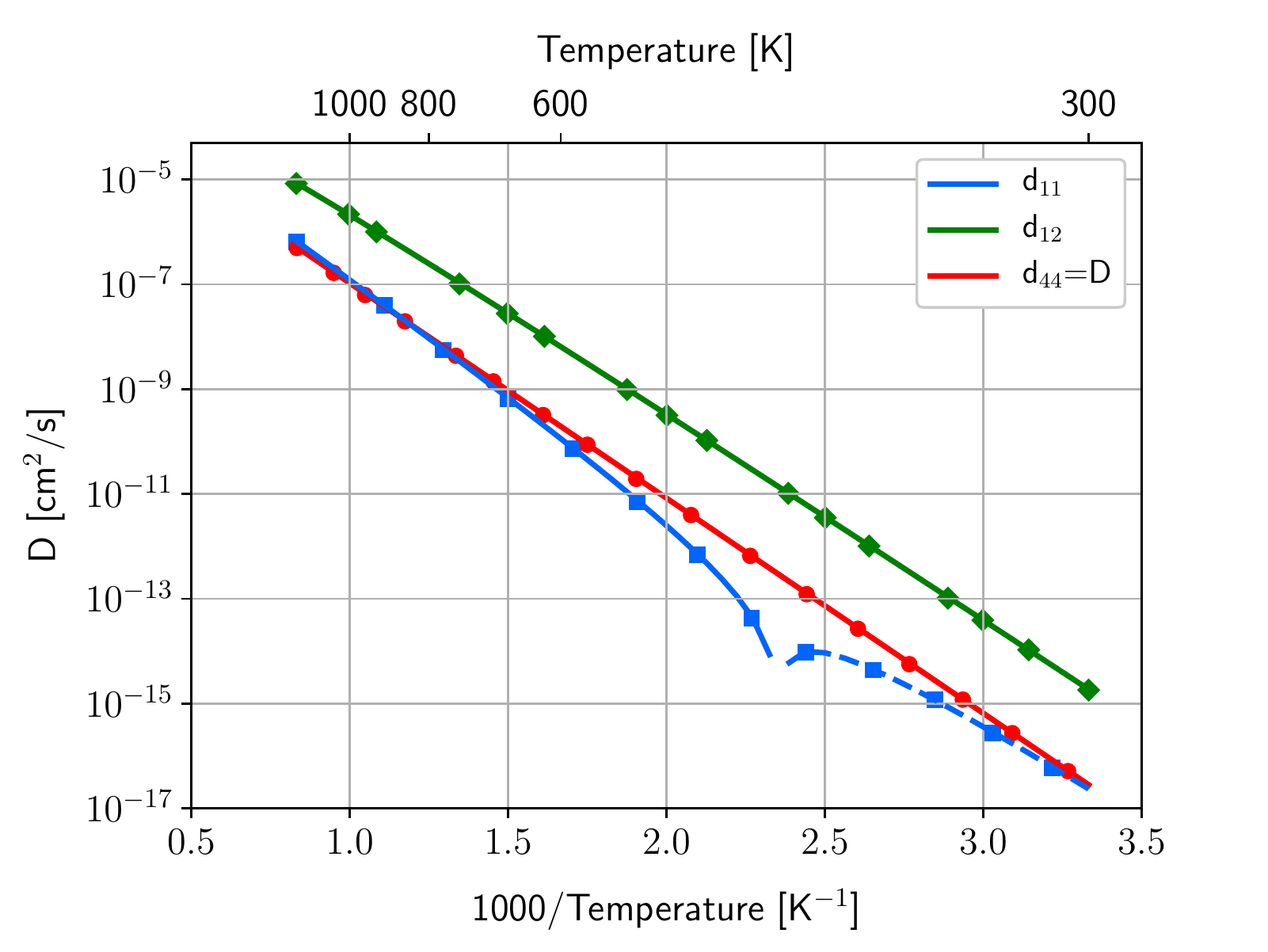}\caption{\label{fig:fec_strain}Components of the elasto-diffusion tensor for interstitial carbon diffusion in BCC iron, compared with the Green function method (markers) \cite{trinkle:2016}. In perfect agreement with the reference work, the $d_{11}$ coefficient shows a sign change at T = 425.1 K (negative values are marked with a dashed line), and the $d_{44}$ coefficient is exactly equal to the carbon diffusivity in the unstrained crystal. }
\end{figure}

KineCluE compares well also with the elasto-diffusion calculations performed with the Green Function (GF) method for interstitial carbon diffusion in BCC iron \cite{trinkle:2016}. Figure \ref{fig:fec_strain} shows the non-null components of the elasto-diffusion tensor $d_{xy} = \mathrm{d}D_{xy}/\mathrm{d}\varepsilon$, obtained in KineCluE by finite differences: $d=(L(\varepsilon_1)-L(\varepsilon_0))/(\varepsilon_1-\varepsilon_0)$, with $\varepsilon_0=0$ and $\varepsilon_1=10^{-6}$. The change of sign of the $d_{11}$ coefficient at $T=425.1$ K is in perfect agreement with the GF calculations; this has been achieved by setting a sufficiently small $\Delta\varepsilon$. Given that the code is already capable of computing derivatives of the transport coefficients (used for instance in the sensitivity study, see Section \ref{sec:sensitivity}), it will be possible in future implementations to obtain the exact value of the derivative $\mathrm{d}D_{xy}/\mathrm{d}\varepsilon$, without the need for finite differences.

\section{Conclusion}
	We introduced the KineCluE code that computes cluster transport coefficients from atomic-scale jump frequencies. The code is highly versatile in terms of crystallographic structures, chemical species, defect types, interaction ranges, and jump mechanisms, which allows for the treatment of a wide range of systems. It is an important tool to bridge two gaps: a "scale" gap and a "concentration" gap. For the former, the cluster expansion of the Onsager matrix enables for an efficient evaluation of kinetic properties as functions of cluster concentrations, since cluster transport coefficients only have to be computed once. Coupling with cluster-dynamics and phase-field methods is thus straightforward, and KineCluE allows for a proper linking of these methods with atomic-scale information such as energies of configurations and migration barriers under well-controlled approximations. Also, cluster quantities are directly useful to object kinetic Monte Carlo methods. Eventually, KineCluE offers new possibilities for studying micro-structure and defect population evolution over significant time scales informed by atomic-scale data. Regarding the concentration gap, available analytical kinetic models have always been separated into exact models for infinitely dilute systems and mean-field approaches for concentrated models. Starting from a dilute-system framework, KineCluE allows for the treatment of larger clusters with respect to previous works, and these clusters contain part of the physics of concentration effects on transport properties. Thus, it is a first but important step in unifying both dilute and concentrated approaches in a unique formalism that would be able to correctly describe intermediate concentrations (i.e., a few percent) which are the most challenging to compute but also the most commonly found in technical applications.
	
	\section*{Acknowledgments}
We thank J.-L. Bocquet and D.R. Trinkle for fruitful discussions.	
This work was partly carried out within the EUROfusion Consortium with financial support from the EURATOM research and training program 2014-2018, Grant No. 633053 (EUROfusion), and No. 661913 (SOTERIA). The views and opinions expressed herein do not necessarily reflect those of the European Commission.

\appendix
	\section{Relationship between M and $\Lambda$ \label{MLambda_anx}}
    
    This appendix derives a relation between quantities $\boldsymbol{\Lambda}$ and $\mathbf{M}$,
    defined in Eqs. \ref{eq:lambda} and \ref{eq:M}, respectively.
    On the one hand, the components of $\mathbf{M}$ are sums over all possible jumps from a
	given configuration -- corresponding to some effective interaction -- of
	thermodynamically averaged jump frequencies. On the other hand, the components of $\boldsymbol{\Lambda}$
	are sums over all possible effective interactions for a given jump.
	We now derive the relation between coefficients $\mathbf{M}_{\sigma_{0}\alpha}$ and
	$\boldsymbol{\Lambda}_{\mu,\alpha\sigma}$, where $\alpha$ is an atom
	or defect species, $\sigma$ represents an effective interaction class, 
    and $\sigma_{0}$ one particular instance of this class. 
	
	Taking a class of effective interactions $\sigma$, all of its instances
	give the same contribution to the $\mathbf{M}$ matrix (in absolute
	value, as the sign might differ). Assuming that there are $N_{\sigma}$
	of such instances per site, the $\mathbf{M}_{\sigma_{0}\alpha}$ components are summed over the ones having a positive contribution, and the ones having
	a negative contribution:
	
	\begin{align}
	N_{\sigma}\mathbf{M}_{\sigma_{0}\alpha}=\sum_{\sigma_{0}\in\sigma_{+}}\mathbf{M}_{\sigma_{0}\alpha}-\sum_{\sigma_{0}\in\sigma_{-}}\mathbf{M}_{\sigma_{0}\alpha}.\label{eq:Mlambda_1}
	\end{align}
	The projection of the jump distance along the CPG direction can also have a positive or negative contribution, such that the sum over sites $s$ in Eq. \ref{eq:M} is divided in two parts:
	
	\begin{align}
	N_{\sigma}\mathbf{M}_{\sigma_{0}\alpha} & =\sum_{j}\left(\sum_{\sigma_{0}\in\sigma_{+}}\sum_{s\in\theta_{j+}^{\alpha}}\chi\left|d_{js}^{\mu}\right|-\sum_{\sigma_{0}\in\sigma_{+}}\sum_{s\in\theta_{j-}^{\alpha}}\chi\left|d_{js}^{\mu}\right|\right.\nonumber \\
	& \left.-\sum_{\sigma_{0}\in\sigma_{-}}\sum_{s\in\theta_{j+}^{\alpha}}\chi\left|d_{js}^{\mu}\right|+\sum_{\sigma_{0}\in\sigma_{-}}\sum_{s\in\theta_{j-}^{\alpha}}\chi\left|d_{js}^{\mu}\right|\right),\label{eq:Mlambda_2}
	\end{align}
	where $\chi=\left\langle \left|n_{\sigma_{0}}\right|n_{j}^{\alpha}m_{js}^{\alpha}\omega_{js}^{\alpha}\right\rangle $.

	In Eq. \ref{eq:lambda}, the flux has been computed in a given direction $d$. The absolute value
	of the reverse flux is identical because of translational invariance.
	It follows that:
	\begin{align}
	2\boldsymbol{\Lambda}_{d,\alpha\sigma} & =\sum_{j}\left(\sum_{s\in\theta_{j+}^{\alpha}}\left|d_{js}^{d}\right|\left\langle \left(n_{\sigma}-\tilde{n}_{\sigma}\right)n_{j}^{\alpha}m_{js}^{\alpha}\omega_{js}^{\alpha}\right\rangle \right.\nonumber \\
	& -\left.\sum_{s\in\theta_{j-}^{\alpha}}\left|d_{js}^{d}\right|\left\langle \left(n_{\sigma}-\tilde{n}_{\sigma}\right)n_{j}^{\alpha}m_{js}^{\alpha}\omega_{js}^{\alpha}\right\rangle \right),\label{eq:Mlambda_3}
	\end{align}
	where the final configuration of a given jump is the initial configuration of the reverse jump, $d_{js}^{d}\left\langle \tilde{n}_{\sigma}n_{j}^{\alpha}m_{js}^{\alpha}\omega_{js}^{\alpha}\right\rangle =-d_{sj}^{d}\left\langle n_{\sigma}n_{s}^{\alpha}m_{sj}^{\alpha}\omega_{sj}^{\alpha}\right\rangle$, and the minus sign comes from the fact that a given jump necessarily has a contribution to $\boldsymbol{\Lambda}$ that is opposite to that of the reverse jump. Using translational invariance, we obtain:
    
    \begin{equation}
    \sum_{j}\sum_{s\in\theta_{j+}^{\alpha}}d_{sj}^{d}\left\langle n_{\sigma}n_{s}^{\alpha}m_{sj}^{\alpha}\omega_{sj}^{\alpha}\right\rangle = \sum_{j}\sum_{s\in\theta_{j-}^{\alpha}}d_{js}^{d}\left\langle n_{\sigma}n_{j}^{\alpha}m_{js}^{\alpha}\omega_{js}^{\alpha}\right\rangle. \label{eq:Mlambda_4}
    \end{equation}
   Hence, Eq. \ref{eq:Mlambda_3} becomes:
   \begin{align}
   2\boldsymbol{\Lambda}_{d,\alpha\sigma} & =\sum_{j}\left(2\sum_{s\in\theta_{j+}^{\alpha}}\left|d_{js}^{d}\right|\left\langle n_{\sigma}n_{j}^{\alpha}m_{js}^{\alpha}\omega_{js}^{\alpha}\right\rangle \right.\nonumber \\
	& -\left.2\sum_{s\in\theta_{j-}^{\alpha}}\left|d_{js}^{d}\right|\left\langle n_{\sigma}n_{j}^{\alpha}m_{js}^{\alpha}\omega_{js}^{\alpha}\right\rangle \right).\label{eq:Mlambda_45}
   \end{align}
    
	The last step is to explicit the sum over instances of the effective interaction class $\sigma$, which is hidden in the thermal average $\left\langle.\right\rangle$. Likewise as previously, this sum is separated into positive and negative contributions:
	
	\begin{align}
	\boldsymbol{\Lambda}_{d,\alpha\sigma} & =\sum_{j}\left(\sum_{\sigma_{0}\in\sigma_{+}}\sum_{s\in\theta_{j+}^{\alpha}}\chi\left|d_{js}^{d}\right|-\sum_{\sigma_{0}\in\sigma_{+}}\sum_{s\in\theta_{j-}^{\alpha}}\chi\left|d_{js}^{d}\right|\right.\nonumber \\
	& \left.-\sum_{\sigma_{0}\in\sigma_{-}}\sum_{s\in\theta_{j+}^{\alpha}}\chi\left|d_{js}^{d}\right|+\sum_{\sigma_{0}\in\sigma_{-}}\sum_{s\in\theta_{j-}^{\alpha}}\chi\left|d_{js}^{d}\right|\right).\label{eq:Mlambda_5}
	\end{align}
		
	Comparing Eqs. \ref{eq:Mlambda_2}  and \ref{eq:Mlambda_5}, $\boldsymbol{\Lambda}_{d,\alpha\sigma}$ and
	$\mathbf{M}_{\sigma_{0}\alpha}$ have the same structure. 
    They differ	by a factor $N_{\sigma}$, which is the number of instances in a given
	effective interaction class, and by the direction along which the jump
	vector is projected: $d_{js}^{d}$ and $d_{js}^{\mu}$, respectively.
	Taking the diffusion direction as the CPG direction, we can write:
    
    \begin{equation}
	N_{\sigma}\mathbf{M}_{\sigma_{0}\alpha}=\boldsymbol{\Lambda}_{\mu,\alpha\sigma}.\label{eq:EQ_M_Lambda_bis}
	\end{equation}
		   
    \section{Site interactions contributions \label{site_anx}}
    
    A configuration is called \textit{associated} when it is located in the sub-space of the configuration space that defines the cluster, and is called \textit{dissociated} in the opposite case. As described in Sec. \ref{sec:siteinter}, only association jumps (i.e., jumps from a dissociated to an associated configuration) may cause sub-cluster configuration probabilities to be non-stationary when sub-cluster interactions are not explicitly accounted for. This is because in our space-exploration method, we only study the transitions from associated configurations, so association jumps are not considered, which may become a problem depending on the symmetry properties of the system. In other words, this issue is related to the definition of the cluster boundary. The idea to build the correct equations for site interactions is to  virtually expand the boundary of the cluster including all the dissociated configurations that are one jump away from a cluster configuration, to make sure that all association jumps are considered. The $\mathbf{\tilde{T}}$ matrix would then become:

\begin{equation}
\left(\begin{array}{cc}
\mathbf{\tilde{T}} & \mathbf{\tilde{T}_{d}}\\
\mathbf{\tilde{T}_{d}}^{t} & \mathbf{\tilde{T}_{\boldsymbol{\delta}}}
\end{array}\right),\label{eq:site0}
\end{equation}
where $\mathbf{\tilde{T}_{d}}$ and $\mathbf{\tilde{T}_{\boldsymbol{\delta}}}$ contain jump frequencies between associated and dissociated configurations only. Let us now define $\mathbf{S}$ and $\mathbf{S_{d}}$ matrices, whose elements indicate the type of site interactions that are included in each associated and dissociated configurations, respectively. Hence, the number of site interactions appearing in the time-derivative of each $n_c$-body and site interaction average is given by:

\begin{equation}
\left(\begin{array}{cc}
\mathbf{\tilde{T}} & \mathbf{\tilde{T}_{d}}\\
\mathbf{\tilde{T}}_{\mathbf{d}}^{t} & \mathbf{\tilde{T}_{\boldsymbol{\delta}}}
\end{array}\right)\left(\begin{array}{c}
\mathbf{S}\\
\mathbf{S_{d}}
\end{array}\right)=\left(\begin{array}{c}
\mathbf{\tilde{T}S}+\mathbf{\tilde{T}_{d}S_{d}}\\
\mathbf{\tilde{T}}_{\mathbf{d}}^{t}\mathbf{S}+\mathbf{\tilde{T}_{\boldsymbol{\delta}}S_{d}}
\end{array}\right).\label{eq:site1}
\end{equation}
The first component of the right-hand side vector is the number of site interactions included in each $n_c$-body effective interaction time-derivative probability, with a contribution from associated ($\mathbf{\tilde{T}S}$) and dissociated ($\mathbf{\tilde{T}_{d}S_{d}}$) configurations. The equations for site interactions are obtained by summing over all $n_c$-body and site effective-interaction contributions:

\begin{equation}
\left(\begin{array}{cc}
\mathbf{S}^{t} & \mathbf{S}_{\mathbf{d}}^{t}\end{array}\right)\left(\begin{array}{c}
\mathbf{\tilde{T}S}+\mathbf{\tilde{T}_{d}S_{d}}\\
\mathbf{\tilde{T}}_{\mathbf{d}}^{t}\mathbf{S}+\mathbf{\tilde{T}_{\boldsymbol{\delta}}S_{d}}
\end{array}\right)=\mathbf{K}+\boldsymbol{\mathbf{S}_{\mathbf{d}}^{t}\mathbf{\tilde{T}_{\boldsymbol{\delta}}S_{d}}},\label{eq:site2}
\end{equation}
where $\mathbf{K}=\mathbf{S}^{t}\mathbf{\tilde{T}S}+\mathbf{S}^{t}\mathbf{D}+\mathbf{D}{}^{t}\mathbf{S}$ and $\mathbf{D}=\mathbf{\tilde{T}}_{\mathbf{d}}\mathbf{S_{d}}$. From Eqs. \ref{eq:site1} and \ref{eq:site2} we can build the matrix version of our problem involving both $n_c$-body effective interactions $\nu$ and effective site interactions $\delta$:

\begin{equation}
\left(\begin{array}{cc}
\tilde{\mathbf{T}} & \mathbf{\tilde{T}S}+\mathbf{D}\\
\mathbf{S}^{t}\mathbf{\tilde{T}}+\mathbf{D}^{t} & \mathbf{K}+\mathbf{S}_{\mathbf{d}}^{t}\mathbf{\tilde{T}_{\boldsymbol{\delta}}S_{d}}
\end{array}\right)\left(\begin{array}{c}
\boldsymbol{\nu}\\
\boldsymbol{\delta}
\end{array}\right)=\left(\begin{array}{c}
\boldsymbol{\Lambda}_{\mu}^{t}\\
\mathbf{S}^{t}\boldsymbol{\Lambda}_{\mu}^{t}+\boldsymbol{\lambda}_{\mu}^{t}
\end{array}\right).\label{eq:site3}
\end{equation}
This linear system that we must solve to obtain the value of effective interactions (site and $n_c$-body) replaces Eq. \ref{eq:systemnosite} ($n_c$-body interactions only). To obtain Eq. \ref{eq:site3}, we added the site interaction contribution to each $n_c$-body effective interaction equation from Eq. \ref{eq:site1}, and the sum of the site interaction contributions are computed from Eq. \ref{eq:site2}. In the right-hand side vector, the $\boldsymbol{\lambda}_{\mu}^{t}$ term accounts for the association jumps (reverse of dissociation jumps appearing in $\boldsymbol{\Lambda}_{\mu}^{t}$).

The next step is to solve the linear system of equations \ref{eq:site3}. First, we express $\nu$ as a function of $\delta$:

\begin{equation}
\boldsymbol{\nu}=\mathbf{\tilde{T}}^{-1}\boldsymbol{\Lambda}_{\mu}^{t}-\left(\mathbf{S}+\tilde{\mathbf{T}}^{-1}\mathbf{D}\right)\boldsymbol{\delta}\label{eq:site4}.
\end{equation}
Then, we write the second (matrix) system of equations \ref{eq:site3}, replacing the value of $\nu$ by Eq. \ref{eq:site4}, which gives access to the value of $\delta$:

\begin{equation}
\boldsymbol{\delta}=\left(\mathbf{S}_{\mathbf{d}}^{t}\mathbf{\tilde{T}_{\boldsymbol{\delta}}S_{d}}-\mathbf{D}^{t}\mathbf{\tilde{T}}^{-1}\mathbf{D}\right)^{-1}\left(\boldsymbol{\lambda}_{\mu}^{t}-\mathbf{D}^{t}\mathbf{\tilde{T}}^{-1}\boldsymbol{\Lambda}_{\mu}^{t}\right).\label{eq:site6}
\end{equation}

The flux vector $\mathbf{J}_{d}=-\frac{1}{V}\left(\boldsymbol{\Lambda}^{0}_{d}-\boldsymbol{\Lambda}_{d}\boldsymbol{\nu
}\right)\boldsymbol{\mu}$ is given below Eq. \ref{eq:T}. In order to take into account site interactions, the correlated contribution to the transport coefficient ($\boldsymbol{\Lambda}_{d}\boldsymbol{\nu}$) is replaced by the following definition:

\begin{equation}
\boldsymbol{\Lambda}_{d}\boldsymbol{\nu}+\left(\boldsymbol{\Lambda}_{d}\mathbf{S}+\boldsymbol{\lambda}_{d}\right)\boldsymbol{\delta}=\boldsymbol{\Lambda}_{d}\mathbf{\tilde{T}}^{-1}\boldsymbol{\Lambda}_{\mu}^{t}+\left(\boldsymbol{\lambda}_{d}-\boldsymbol{\Lambda}_{d}\mathbf{\tilde{T}}^{-1}\mathbf{D}\right)\boldsymbol{\delta}.\label{eq:site7}
\end{equation}
Finally, we obtain the expression for the cluster transport coefficients that takes into account site interactions:

\begin{equation}
\mathbf{L}_{d}^{\mathrm{eq}}\left(c\right)=\frac{\boldsymbol{\Lambda}_{d}^{0}-\left[\boldsymbol{\Lambda}_{d}\mathbf{\tilde{\mathbf{T}}^{-1}}\boldsymbol{\Lambda}_{\mu}^{t}+\boldsymbol{\gamma}_{d}\boldsymbol{\tau}^{-1}\boldsymbol{\gamma}_{\mu}^{t}\right]}{V},\label{eq:site8}
\end{equation}
with $\boldsymbol{\gamma}_{d}=\boldsymbol{\lambda}_{d}-\boldsymbol{\Lambda}_{d}\mathbf{\tilde{T}}^{-1}\mathbf{D}$ and $\boldsymbol{\tau}=\mathbf{S}_{\mathbf{d}}^{t}\mathbf{\tilde{T}_{\boldsymbol{\delta}}S_{d}}-\mathbf{D}^{t}\mathbf{\tilde{T}}^{-1}\mathbf{D}$. The site-interaction contribution appears as a separate term in Eq. \ref{eq:site8}, namely $\boldsymbol{\gamma}_{d}\boldsymbol{\tau}^{-1}\boldsymbol{\gamma}_{\mu}^{t}$. It is interesting to note that because $\mathbf{\tilde{T}_{\boldsymbol{\delta}}}$ is symmetric, the site-interaction contribution is also necessarily symmetric -- at least for diffusion along $\vec{e}_{\mu}$ -- and the cluster transport-coefficient matrix remains symmetric. Accounting for this correction requires the computation of three quantities: $\mathbf{D}$, $\boldsymbol{\lambda}_{d}$ and $\mathbf{S}_{\mathbf{d}}^{t}\mathbf{\tilde{T}_{\boldsymbol{\delta}}S_{d}}$ (see Eqs. \ref{eq:msite4}-\ref{eq:msite6}), all being obtained "on the fly" as we compute $\mathbf{\tilde{T}}$, with only a small additional computational cost.
        
	\section{Sensitivity study \label{sensi_anx}}
	
	Transport coefficients for clusters of more than two components depend
	on a large number of jump frequencies, and this number increases exponentially
	with the number of cluster components. The sensitivity study
	is a KineCluE routine designed to identify the most important jump
	frequencies by computing the local derivatives of cluster transport
	coefficients with respect to the jump frequencies. Starting from a set
	of jump frequencies -- which might not be very accurate -- this routine
	identifies the most critical jump frequencies, i.e., the ones transport coefficients are most sensitive to, and must therefore be accurately computed.
	Because the method is local, several iterations might be necessary
	to identify the full set of critical jump frequencies. 
	
	Let us assume that we have a set of jump frequencies $\mathbf{W}_{0}=\left(\omega_{i}^{0}\right)$;
	we are then able to compute a numerical value for $\mathbf{L}_{d}\left(\mathbf{W}_{0}\right)$.
	Now we want to know how $\mathbf{L}_{d}\left(\mathbf{W}_{0}\right)$ would change if some of the jump frequencies $\omega_{i}$ would be modified. To this end, we
	will compute the gradient of $\mathbf{L}_{d}$ in the jump-frequency phase
	space which points towards the direction of the highest change in
	$\mathbf{L}_{d}$: 
	
	\begin{equation}
	\boldsymbol{\nabla}\mathbf{L}_{d}=\left(\left.\frac{\partial\mathbf{L}_{d}}{\partial\omega_{i}}\right|_{\mathbf{W}_{0}}\right).\label{eq:gradL}
	\end{equation}
	We compute the partial derivatives using the matrix form of cluster transport
	coefficients: $\mathbf{L}_{d}\left(c\right)=\boldsymbol{\Lambda}^{0}_{d}-\boldsymbol{\Lambda}_{d}\tilde{\mathbf{T}}^{-1}\boldsymbol{\Lambda}_{\mu}^{t}$ (Eq. \ref{eq:matrixnotation}):
	
	\begin{equation}
	\frac{\partial\mathbf{L}_{d}}{\partial\omega_{i}}=\frac{\partial\boldsymbol{\Lambda}^{0}_{d}}{\partial\omega_{i}}-\frac{\partial\boldsymbol{\Lambda}_{d}}{\partial\omega_{i}}\tilde{\boldsymbol{T}}^{-1}\boldsymbol{\Lambda}_{\mu}^{t}-\boldsymbol{\Lambda}_{d}\frac{\partial\tilde{\mathbf{T}}^{-1}}{\partial\omega_{i}}\boldsymbol{\Lambda}_{\mu}^{t}-\boldsymbol{\Lambda}_{d}\tilde{\mathbf{T}}^{-1}\frac{\partial\boldsymbol{\Lambda}_{\mu}^{t}}{\partial\omega_{i}}.\label{eq:deriv1}
	\end{equation}
	
	Note that all the above matrices and derivatives are evaluated at
	$\mathbf{W}_{0}$. The terms in matrices $\boldsymbol{\Lambda}^{0}$,
	$\boldsymbol{\Lambda}$ and $\tilde{\mathbf{T}}$ are linear
	combinations of jump frequencies, and we have computed their analytical
	form in the analytical part of the code, so their derivatives
	with respect to a given jump frequency $\omega_{i}$ are straightforward
	to compute. The remaining issue is that we do not have the analytical
	expression of $\tilde{\mathbf{T}}^{-1}$, and the latter can be prohibitively long
	to compute because $\tilde{\mathbf{T}}$ is a square matrix of size equal to the number
	of effective interactions in the system. Thus, we use the following
	identity:
	\begin{align}
	\tilde{\mathbf{T}}\tilde{\mathbf{T}}^{-1}=\mathbf{I} & \Rightarrow\;\frac{\partial\tilde{\mathbf{T}}}{\partial\omega_{i}}\tilde{\mathbf{T}}^{-1}+\tilde{\mathbf{T}}\frac{\partial\tilde{\mathbf{T}}^{-1}}{\partial\omega_{i}}=\mathbf{0}\nonumber \\
	& \Rightarrow\;\frac{\partial\tilde{\mathbf{T}}^{-1}}{\partial\omega_{i}}=-\tilde{\mathbf{T}}^{-1}\frac{\partial\tilde{\mathbf{T}}}{\partial\omega_{i}}\tilde{\mathbf{T}}^{-1}.\label{eq:idinvT}
	\end{align}
	Inserting Eq. \ref{eq:idinvT} in Eq. \ref{eq:deriv1} amounts to
	a practically convenient expression:
	\begin{align}
	\frac{\partial\mathbf{L}_{d}}{\partial\omega_{i}}= & \frac{\partial\boldsymbol{\Lambda}^{0}_{d}}{\partial\omega_{i}}-\frac{\partial\boldsymbol{\Lambda}_{d}}{\partial\omega_{i}}\tilde{\mathbf{T}}^{-1}\boldsymbol{\Lambda}_{\mu}+ \boldsymbol{\Lambda}_{d}\tilde{\mathbf{T}}^{-1}\frac{\partial\tilde{\mathbf{T}}}{\partial\omega_{i}}\tilde{\mathbf{T}}^{-1}\boldsymbol{\Lambda}_{\mu} \nonumber \\
	& - \boldsymbol{\Lambda}_{d}\tilde{\mathbf{T}}^{-1}\frac{\partial\boldsymbol{\Lambda}_{\mu}}{\partial\omega_{i}}.\label{eq:practical}
	\end{align}
	Now all the derivatives can be easily computed from the analytical
	expressions of $\boldsymbol{\Lambda}^{0}$, $\boldsymbol{\Lambda}$ and
	$\tilde{\mathbf{T}}$. Then, for a given set of jump frequencies
	$\mathbf{W}_{0}$, all terms located in Eq. \ref{eq:practical} are
	directly computed numerically, and the numerical value of $\tilde{\mathbf{T}}^{-1}$
	is already known from the calculation of $\mathbf{L}_{d}\left(\mathbf{W}_{0}\right)$.
	The direction $\mathbf{V}$ in jump frequency space along which $\mathbf{L}_{d}$
	varies the most around $\mathbf{W}_{0}$ is given by the normalized gradient:
	
	\begin{equation}
	\mathbf{V}=\frac{\nabla\mathbf{L}_{d}}{\left\Vert \nabla\mathbf{L}_{d}\right\Vert }=\frac{\nabla\mathbf{L}_{d}}{\sqrt{\nabla\mathbf{L}_{d}.\nabla\mathbf{L}_{d}}}.\label{eq:direction}
	\end{equation}
	Practically speaking, vector $\mathbf{V}$  gives the weighted combination of
	jump frequencies that is able to affect $\mathbf{L}_{d}\left(\mathbf{W}_{0}\right)$ the most.
	These weights then evidence the most critical jump frequencies
	to get an accurate estimation of $\mathbf{L}_{d}$ around $\mathbf{W}_{0}$.
	These jump frequencies should be computed accurately (e.g.,
	using first-principles calculations) to obtain a new estimation of
	$\mathbf{L}_{d}$ around a new point in the jump-frequency space, $\mathbf{W}_{1}$.
	Then again, the corresponding $\mathbf{V}$ vector is obtained to verify which
	are the most relevant jump frequencies around that point, and so on
	and so forth. 
	
	The remaining question is: when should we stop? Since the analysis is  local, when we change some of the jump frequencies, we will not get the same set of critical jump frequencies, and eventually
	we might end up computing all of them, which is what we wanted to avoid to start with. The idea is to take a set of jump frequencies among the ones with the largest $\mathbf{V}$ vector components and compute the numerical value of transport coefficients for changes in these jump frequency values. A batch calculation feature is provided in KineCluE for this purpose. Then, one can decide if it is worth computing the value of a given saddle-point configuration precisely, or if it does not affect the cluster transport coefficients within a tolerance chosen by the user. An example of such procedure is presented in Sec. \ref{sec:sensitivity}.
	
	When site interactions are explicitly taken into account (see Sec. \ref{sec:siteinter}), Eq. \ref{eq:practical} comes with additional terms, but the computation remains straightforward because all the terms in matrices $\mathbf{D}$, $\boldsymbol{\lambda}_d$ and $\mathbf{S}_{\mathbf{d}}^{t}\mathbf{\tilde{T}_{\boldsymbol{\delta}}S_{d}}$ are linear combinations of jump frequencies:
	
		\begin{align}
	& \frac{\partial\mathbf{L}_{d}}{\partial\omega_{i}}=  \frac{\partial\boldsymbol{\Lambda}^{0}_{d}}{\partial\omega_{i}}
	-\frac{\partial\boldsymbol{\Lambda}_{d}}{\partial\omega_{i}}\tilde{\mathbf{T}}^{-1}\boldsymbol{\Lambda}_{\mu} 
	-\frac{\partial\boldsymbol{\gamma}_{d}}{\partial\omega_{i}}\tau^{-1}\boldsymbol{\gamma}_{\mu}^{t}\nonumber \\
	& + \boldsymbol{\Lambda}_{d}\tilde{\mathbf{T}}^{-1}\left(\frac{\partial\tilde{\mathbf{T}}}{\partial\omega_{i}}\tilde{\mathbf{T}}^{-1}\boldsymbol{\Lambda}_{\mu}-\frac{\partial\boldsymbol{\Lambda}_{\mu}}{\partial\omega_{i}}\right) \nonumber \\
	& + \boldsymbol{\gamma}_{d}\tau^{-1}\left(\frac{\partial\boldsymbol{\tau}}{\partial\omega_{i}}\boldsymbol{\tau}^{-1}\boldsymbol{\gamma}_{\mu}^{t}-\frac{\partial\boldsymbol{\gamma}_{\mu}^{t}}{\partial\omega_{i}}\right),\label{eq:practical2}
	    \end{align}
	with:
	
	\begin{align}
	    & \frac{\partial\boldsymbol{\tau}}{\partial\omega_{i}}=\frac{\partial\left(\mathbf{S}_{\mathbf{d}}^{t}\mathbf{\tilde{T}_{\boldsymbol{\delta}}S_{d}}\right)}{\partial\omega_{i}}  -\frac{\partial\mathbf{D}^{t}}{\partial\omega_{i}}\tilde{\mathbf{T}}^{-1}\mathbf{D} \nonumber \\
	    & +\mathbf{D}^{t}\tilde{\mathbf{T}}^{-1}\left(\frac{\partial\tilde{\mathbf{T}}}{\partial\omega_{i}}\tilde{\mathbf{T}}^{-1}\mathbf{D}-\frac{\partial\mathbf{D}}{\partial\omega_{i}}\right), \label{eq:sitesensi1}
	\end{align}
	and:
	\begin{align}
	    \frac{\partial\boldsymbol{\gamma}_{d}}{\partial\omega_{i}}=\frac{\partial\boldsymbol{\lambda}_{d}}{\partial\omega_{i}}+\boldsymbol{\Lambda}_{d}\tilde{\mathbf{T}}^{-1}\left(\frac{\partial\tilde{\mathbf{T}}}{\partial\omega_{i}}\tilde{\mathbf{T}}^{-1}\mathbf{D}-\frac{\partial\mathbf{D}}{\partial\omega_{i}}\right)-\frac{\partial\boldsymbol{\Lambda}_{d}}{\partial\omega_{i}}\tilde{\mathbf{T}}^{-1}\mathbf{D}. \label{eq:sitesensi2}
	\end{align}
	
	Another possible approach, which is not yet part of the KineCluE code, is to compute the Taylor expansion of transport coefficients around a reference point in the jump-frequency space. This may be numerically advantageous because the cluster transport coefficient derivative of any order uses the same derivatives as the gradient, so no additional analytical effort is needed because it would be only a numerical application requiring matrix multiplications. The $N^\mathrm{th}$-order Taylor expansion for multivariate functions reads:
	
	\begin{equation}
	\ensuremath{\mathbf{L}_{d}}=\sum_{|n|=0}^{N}\prod_{i=0}^{n}\left[\dfrac{\left(\omega_{i}-\omega_{i}^{0}\right)^{n_{i}}}{n_{i}!}\left.\dfrac{\partial^{n_{i}}\ensuremath{\mathbf{L}_{d}}}{\partial\omega_{i}^{n_{i}}}\right\rfloor _{\ensuremath{\mathbf{W}_{0}}}\right] \label{eq:taylor}
	\end{equation}
	where $|n|=k$ denotes a sum over all possible values of $\left\lbrace n_i\right\rbrace$ where $\sum_i n_i=k$. The $n^{th}$ order derivative (for $n>1$) of $\mathbf{L}_{d}$ with respect to $n$ jump frequencies (eventually some jump frequencies may appear more than once) is expressed as  (we do not take into account site interactions for simplicity):: 
	
	\begin{align}
	& -\dfrac{\partial^{n}\mathbf{L}_{d}}{\partial\omega^{n}}=\boldsymbol{\Lambda}_{d}\dfrac{\partial^{n}\tilde{\mathbf{T}}^{-1}}{\partial\omega^{n}}\boldsymbol{\Lambda}_{\mu}^{t} \nonumber\\
	& +\sum_{\gamma=1}^{n}\left[\dfrac{\partial\boldsymbol{\Lambda}_{d}}{\partial\omega_{\gamma}}\dfrac{\partial^{n-1}\tilde{\mathbf{T}}^{-1}}{\partial\omega_{\neq\gamma}^{n-1}}\boldsymbol{\Lambda}_{\mu}^{t}+\boldsymbol{\Lambda}_{d}\dfrac{\partial^{n-1}\tilde{\mathbf{T}}^{-1}}{\partial\omega_{\neq\gamma}^{n-1}}\dfrac{\partial\boldsymbol{\Lambda}_{\mu}^{t}}{\partial\omega_{\gamma}}\right] \nonumber\\
	& +\sum_{1\leq\gamma<\delta\leq n}\left[\dfrac{\partial\boldsymbol{\Lambda}_{d}}{\partial\omega_{\gamma}}\dfrac{\partial^{n-2}\tilde{\mathbf{T}}^{-1}}{\partial\omega_{\neq\gamma,\delta}^{n-2}}\dfrac{\partial\boldsymbol{\Lambda}_{\mu}^{t}}{\partial\omega_{\delta}}+\dfrac{\partial\boldsymbol{\Lambda}_{d}}{\partial\omega_{\delta}}\dfrac{\partial^{n-2}\tilde{\mathbf{T}}^{-1}}{\partial\omega_{\neq\gamma,\delta}^{n-2}}\dfrac{\partial\boldsymbol{\Lambda}_{\mu}^{t}}{\partial\omega_{\gamma}}\right].
	\label{eq:nth-order}
	\end{align}
	The $n^{th}$ order derivative of the inverse of the $\tilde{\mathbf{T}}$ matrix is given as:
	
	\begin{equation}
	\dfrac{\partial^{n}\tilde{\mathbf{T}}^{-1}}{\partial\omega^{n}}=\left[\sum_{\rho}\prod_{i=1}^{n}\left(-\tilde{\mathbf{T}}^{-1}\dfrac{\partial\tilde{\mathbf{T}}}{\partial\omega_{\rho\left(i\right)}}\right)\right]\tilde{\mathbf{T}}^{-1},
	\label{eq:T-1nthorder}
	\end{equation}
	where the sum over $\rho$ denotes a sum over the possible permutations of jump frequencies and a short-hand notation is used in the two previous equations:
	\begin{equation}
	\dfrac{\partial^{n-1}A}{\partial\omega_{\neq j}^{n-1}}=\dfrac{\partial^{n-1}A}{\partial\omega_{1}...\partial\omega_{j-1}\partial\omega_{j+1}...\partial\omega_{n}}
	\label{eq:notation}
	\end{equation}
	
	This way of computing the local variations of cluster transport coefficients in the jump-frequency space might be more efficient from a numerical point of view than re-computing transport coefficients for each values of the most important jump frequencies. If testing shows that this Taylor expansion formalism is more efficient, it will be implemented in future versions of KineCluE.

\section*{References}
\bibliographystyle{elsarticle-num}
\bibliography{bib}

\end{document}